\documentclass[aps,prd,superscriptaddress,nofootinbib]{revtex4}
\usepackage{amssymb}
\usepackage{amsmath}
\usepackage{epsfig}
\usepackage{subfig}
\usepackage{graphics}

\newcommand{\bx}{\boldsymbol{x}}
\newcommand{\by}{\boldsymbol{y}}
\newcommand{\bz}{\boldsymbol{z}}
\newcommand{\bk}{\boldsymbol{k}}
\newcommand{\bb}{\boldsymbol{b}}
\newcommand{\bc}{\boldsymbol{c}}
\newcommand{\rmi}{\mathrm{i}}
\newcommand{\rmd}{\mathrm{d}}
\newcommand{\rme}{\mathrm{e}}

\allowdisplaybreaks[4]

\begin{document}

\title{Multipole scattering amplitudes in the Color Glass Condensate formalism}

\author{Yu Shi}\email{physhiyu@mails.ccnu.edu.cn}
\affiliation{Key Laboratory of Quark and Lepton Physics (MOE) and Institute of Particle Physics,Central China Normal University, Wuhan 430079, China}

\author{Cheng Zhang}\email{zhangcheng@mails.ccnu.edu.cn}
\affiliation{Key Laboratory of Quark and Lepton Physics (MOE) and Institute of Particle Physics,Central China Normal University, Wuhan 430079, China}

\author{Enke Wang}\email{wangek@mail.ccnu.edu.cn}
\affiliation{Key Laboratory of Quark and Lepton Physics (MOE) and Institute of Particle Physics,Central China Normal University, Wuhan 430079, China}

%\author{Bo-Wen Xiao}\email{bxiao@mail.ccnu.edu.cn}
%\affiliation{Key Laboratory of Quark and Lepton Physics (MOE) and Institute of Particle Physics,Central China Normal University, Wuhan 430079, China}

\begin{abstract}
\vspace{0.cm}
We evaluate the octupole in the large-$N_c$ limit in the McLerran-Venugopalan model, and derive a general expression of the 2n-point correlator, which can be applied in analytical studies of the multi-particle production in the scatterings between hard probes and dense targets.
\end{abstract}
\maketitle

\section{Introduction}

Multi-point functions from the dipole amplitude to 2n-point correlators described by Wilson lines play an important role in calculating the cross sections of the multi-particle production processes, which could take place in deep inelastic scattering (DIS) experiments and proton-nucleus (pA) collisions.

In high-energy DIS and pA collisions, high-energy partons multiply scatter with a small-x gluon field inside a large nucleus in an eikonal way \cite{Kovner:2001vi}. In the Color Glass Condensate (CGC) \cite{Iancu:2002xk,arXiv:1002.0333} formalism, the color source in the heavy target is regarded as a classical color field, and the color charges are assumed to obey the Guassian distribution. In the above framework, multipoles written in terms of Guassian average of Wilson lines appear in the multiple scattering cross sections, and each Wilson line in the scattering amplitudes represents a parton traversing the gluon field with a fixed transverse coordinate in the eikonal approximation, and resums the multiple interactions between the parton and the dense target.

Dipoles and quadrupoles correspond to two types of gluon distributions \cite{Dominguez:2010xd,Catani:1990eg,Collins:1991ty} (the Weisz\"acker-Williams gluon distribution \cite{Kovchegov:1998bi,McLerran:1998nk} and the dipole gluon distribution). The evolution of the dipole gluon distribution obeys the Balitsky-Kovchegov equation \cite{Balitsky:1995ub+X,Kovchegov:1999yj}, while the Weisz\"acker-Williams gluon distribution is governed by the evolution of the quadrupole amplitude, which have been successfully evaluated in Ref. \cite{Dominguez:2011gc} by using the JIMWLK \cite{Jalilian-Marian:1997jx+X,Ferreiro:2001qy} Hamiltonian method. As shown in Ref. \cite{Dominguez:2010xd}, the generalized factorization in high-density QCD can be established in the CGC formalism with only dipole and quadrupole amplitudes. Higher multiple scattering amplitudes, such as 6-point (sextupole) and 8-point (octupole) functions are suppressed in the large-$N_c$ limit in physical processes \cite{Dominguez:2012ad}. Nevertheless, it is interesting to evaluate the leading $N_c$ contribution of higher-point correlators, which may help us to better understand the underlying dynamics and estimate the size of finite-$N_c$ corrections to multi-parton productions. The complexity naturally arises as we compute the correlators consisting of more than four Wilson lines. Fortunately, the large-$N_c$ limit simplifies the calculation and makes it possible to give the relatively simple analytical results for higher-point correlators.

In this paper, we evaluate the correlator of eight Wilson lines and then conjecture a general 2n-point expression, which can be used as the initial condition for small-$x$ evolutions \cite{Iancu:2011ns,Dumitru:2011vk} of any even number point correlators.  We also use $n=3$ case as an example to show how the general formula gives the correct expression for the six-point function.

This paper is organised as follows. In section II, we outline the main aspects of the McLerran-Venugopalan (MV) \cite{McLerran:1993ni} model and briefly review the results of the dipole, quadrupole and sextupole amplitudes. We next proceed to compute the octupole amplitude and conjecture a general expression for 2n-point correlators in section III. Finally section IV is devoted to the conclusion.

\section{Correlators in the McLerran-Venugopalan model}

We aim to compute the 2n-point correlator, namely

\begin{eqnarray}
\frac{1}{N_c}\left\langle\text{Tr}\left[U(\bx_{1\perp})U(\bx_{2\perp})^{\dagger}...U(\bx_{2n-1\perp})U(\bx_{2n\perp})^{\dagger}\right]\right\rangle,
\end{eqnarray}
where in the MV model, $U(\bx_\perp)$ is a Wilson line in the fundamental representation defined as

\begin{equation}
 U(\bx_\perp)=\mathcal{P}\exp\Biggl[-\rmi g^2\int_{-\infty}^{+\infty}\rmd x^+\rmd^2 \bz_\perp G_0(\bx_\perp-\bz_\perp)\rho_a(x^+,\bz_\perp)t^a\Biggr],
\end{equation}
where $t^a$ is a color matrix in the fundamental $\textrm{SU}(N_\textrm{c})$ representation, $\rho_a$ is the color source inside a target, and $\mathcal{P}$ represents $x^+$ ordering operator. $G_0$ is the 2-dimensional propagator which satisfies

\begin{equation}
 \frac{\partial^2}{\partial\bx_\perp^2} G_0(\bx_\perp-\bz_\perp)=\delta(\bx_\perp-\bz_\perp).
\end{equation}

One can solve the above equation and get the explicit form as

\begin{equation}
G_0(\bx_\perp-\bz_\perp)=\int\frac{\rmd^2\bk_\perp}{(2\pi)^2}\frac{\rme^{\rmi \bk_\perp\cdot(\bx_\perp-\bz_\perp)}}{\bk^2_\perp}.
\end{equation}

As mentioned before, our calculation is based on the MV model, which gives the color field average of a given physical quantity as the Guassian weighted functional integral

\begin{equation}
\langle f[\rho]\rangle=\int\mathcal{D}\rho\exp\left\{-\int \rmd^2\bx\,\rmd^2\by\,\rmd z^+\,\frac{\rho_c(z^+,\bx)\rho_c(z^+,\by)}{2\mu^2(z^+)}\right\}f[\rho],
\end{equation}
where $\mu^2(z^+)$ is the variance of the Guassian distribution of color field representing the color charge density at coordinate $z^+$, and its integration over $z^+$ is proportional to the saturation momentum square $Q_{\rm s}^2$.

With the help of the Wick's theorem, any correlators can be obtained by the most elementary correlator which reads

\begin{equation}
\langle\rho_a(x^+,\bx)\rho_b(y^+,\by)\rangle=\delta_{ab}\delta(x^+-y^+)\delta(\bx-\by)\mu^2(x^+).
\end{equation}

Under the framework above, several studies \cite{HiroFujii,Gelis:2001da,Blaizot:2004wv,Dominguez:2012ad,Fukushima:2007dy,Dominguez:2008aa,Dominguez:2011wm,Kovchegov:2008mk,Marquet:2010cf} have obtained the finite-$N_c$ expressions for the dipole and the quadrupole in the MV model, as well as the large-$N_c$ expressions for the quadrupole and the sextupole. In the following discussion, in order to be self-contained, we summarize the known results in the large-$N_c$ limit first. The dipole amplitude reads \cite{Gelis:2001da}

\begin{eqnarray}
\frac{1}{N_c}\left\langle\text{Tr}\left[U_{1}U_{2}^{\dagger}\right]\right\rangle\
=e^{-\Gamma_{12}},
\end{eqnarray}
where

\begin{eqnarray}
&&\Gamma_{ij}=\frac{C_F}{2}\mu^2(L_{ii}+L_{jj}-2L_{ij}),
\\
&&\mu^2=\int\rmd x^{+}\mu^{2}\left(x^{+}\right),\label{2-point}
\end{eqnarray}
with $L_{ij}$ given by the two-dimensional massless propagator $G_0$ as

\begin{eqnarray}
L_{ij}=g^4\int \rmd^2\bz_\perp\;G_0(\bx_{i\perp}-\bz_\perp)G_0(\by_{i\perp}-\bz_\perp).
\end{eqnarray}

The quadrupole can be written as \cite{Blaizot:2004wv,Dominguez:2011wm}

\begin{eqnarray}
&&\frac{1}{N_c}\left\langle\text{Tr}\left[ U_{1}U_{2}^{\dagger }U_{3}U_{4}^{\dagger}\right]\right\rangle\nonumber
\\
&&=e^{-\Gamma_{12}-\Gamma_{34}}-\frac{F_{1234}}{F_{1324}}\left[e^{-\Gamma_{12}-\Gamma_{34}}-e^{-\Gamma_{14}-\Gamma_{32}}\right].\label{4-point}
\end{eqnarray}

The sextupole \cite{Dominguez:2012ad} is given by the following expression in the MV model

\begin{eqnarray}
&&\frac{1}{N_c}\left\langle\text{Tr}\left[ U_{1}U_{2}^{\dagger }U_{3}U_{4}^{\dagger}U_{5}U_{6}^{\dagger }\right]\right\rangle\nonumber
\\
&&=e^{-\Gamma_{12}-\Gamma_{34}-\Gamma_{56}}-\frac{F_{1234}}{F_{1324}}\left[e^{-\Gamma_{12}-\Gamma_{34}}-e^{-\Gamma_{14}-\Gamma_{32}}\right]e^{-\Gamma_{56}}\nonumber
\\
&&-\frac{F_{1256}}{F_{1526}}\left[e^{-\Gamma_{12}-\Gamma_{56}}-e^{-\Gamma_{16}-\Gamma_{52}}\right]e^{-\Gamma_{34}}-\frac{F_{3456}}{F_{3546}}\left[e^{-\Gamma_{34}-\Gamma_{56}}-e^{-\Gamma_{36}-\Gamma_{54}}\right]e^{-\Gamma_{12}}\nonumber
\\
&&+F_{1234}F_{1456}\left[\frac{e^{-\Gamma_{12}-\Gamma_{34}-\Gamma_{56}}}{F_{1324}G}-\frac{e^{-\Gamma_{14}-\Gamma_{32}-\Gamma_{56}}}{F_{1324}F_{1546}}+\frac{e^{-\Gamma_{16}-\Gamma_{32}-\Gamma_{54}}}{F_{1546}G}\right]\nonumber \\
&&+F_{1256}F_{2543}\left[\frac{e^{-\Gamma_{12}-\Gamma_{34}-\Gamma_{56}}}{F_{1526}G}-\frac{e^{-\Gamma_{16}-\Gamma_{34}-\Gamma_{52}}}{F_{1526}F_{2453}}+\frac{e^{-\Gamma_{16}-\Gamma_{32}-\Gamma_{54}}}{F_{2453}G}\right]\nonumber \\
&&+F_{3456}F_{1236}\left[\frac{e^{-\Gamma_{12}-\Gamma_{34}-\Gamma_{56}}}{F_{3546}G}-\frac{e^{-\Gamma_{12}-\Gamma_{36}-\Gamma_{54}}}{F_{3546}F_{1326}}+\frac{e^{-\Gamma_{16}-\Gamma_{32}-\Gamma_{54}}}{F_{1326}G}\right],
\label{6-point}
\end{eqnarray}
with

\begin{eqnarray}
&&F_{ijkl}=L_{ik}-L_{jk}+L_{jl}-L_{il},
\\
&&G=L_{12}+L_{34}+L_{56}-L_{16}-L_{32}-L_{54}.
\end{eqnarray}

From the above three expressions for the dipole, quadupole, and sextupole amplitudes, we find that higher-point functions have terms with a larger number of factors $F_{jikl}$'s, which equals the number of transitions between color singlet states. For example, the sextupole amplitude contains three types of terms, corresponding to singlet states with no color transitions, one and two times of color transitions, respectively.

\section{The evaluation of 2n-point correlators in the large-$N_c$ limit}

%\begin{figure}[!h]
%\vskip0.0\linewidth
%\centerline{
%\includegraphics[width = 0.15\linewidth]{o1.eps}
%\includegraphics[width = 0.15\linewidth]{o2.eps}
%\includegraphics[width = 0.15\linewidth]{o1.eps}
%\includegraphics[width = 0.15\linewidth]{o2.eps}
%\includegraphics[width = 0.15\linewidth]{o1.eps}
%\includegraphics[width = 0.15\linewidth]{o2.eps}
%\includegraphics[width = 0.15\linewidth]{o1.eps}
%}
%\caption{}
%\label{}
%\end{figure}

\begin{figure}[!h]
\centering
\begin{minipage}{2cm}
\includegraphics[width = .8\linewidth]{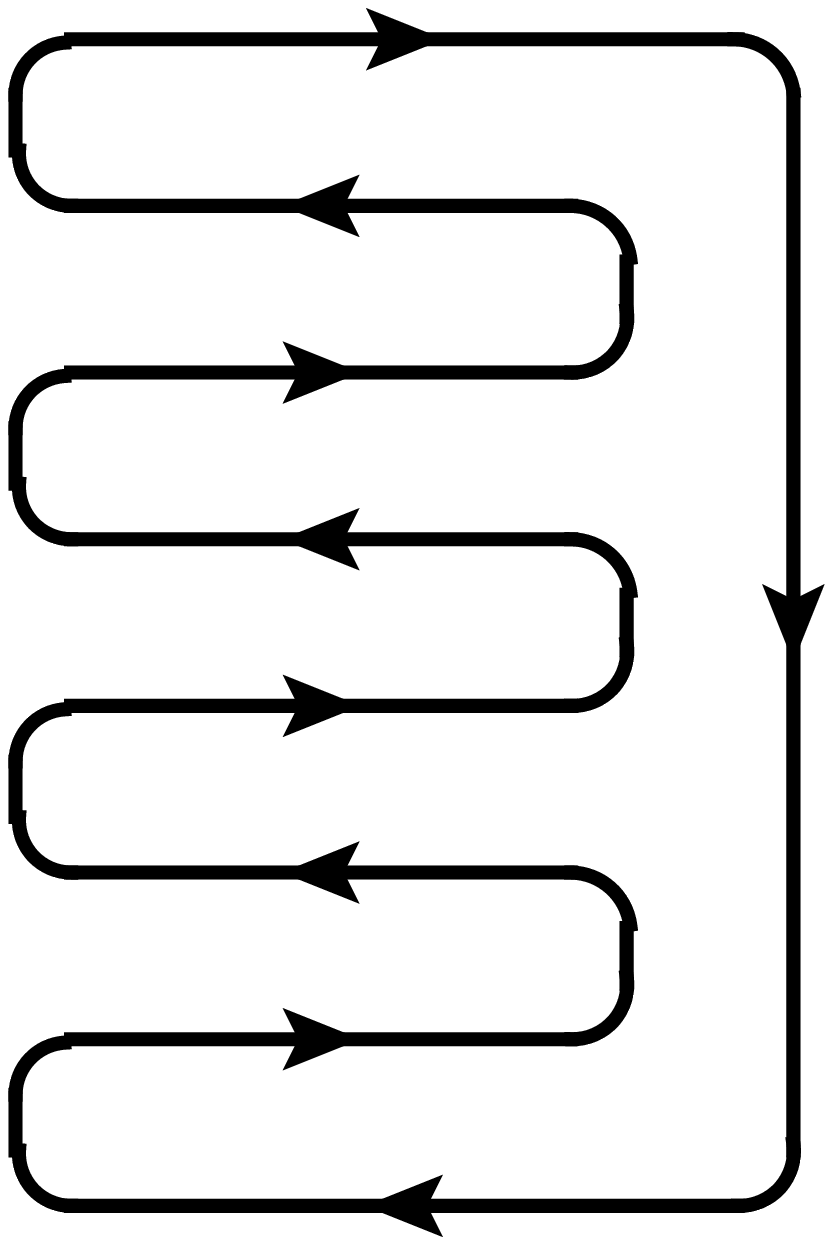}
\hspace{3cm}
\end{minipage}
\begin{minipage}{2cm}
\includegraphics[width = .8\linewidth]{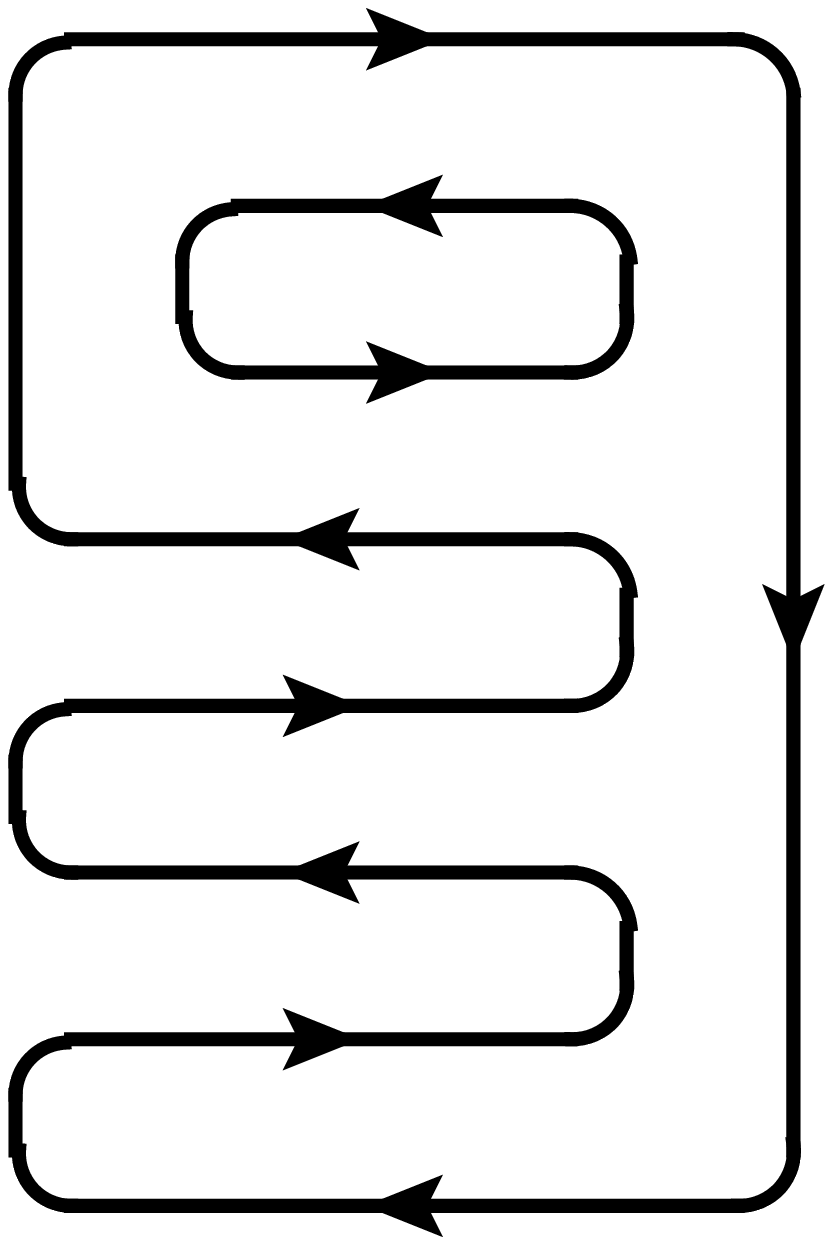}
\hspace{3cm}
\end{minipage}
\begin{minipage}{2cm}
\includegraphics[width = .8\linewidth]{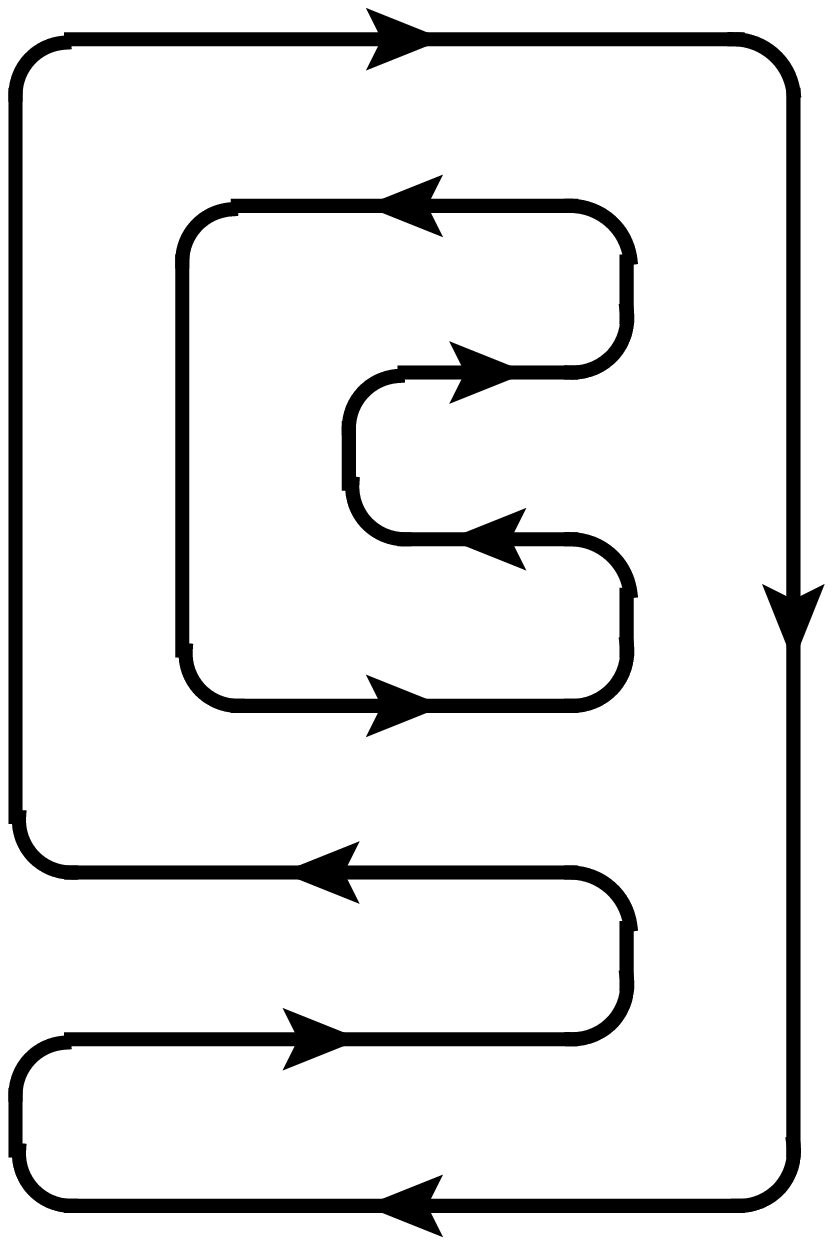}
\hspace{3cm}
\end{minipage}
\begin{minipage}{2cm}
\includegraphics[width = .8\linewidth]{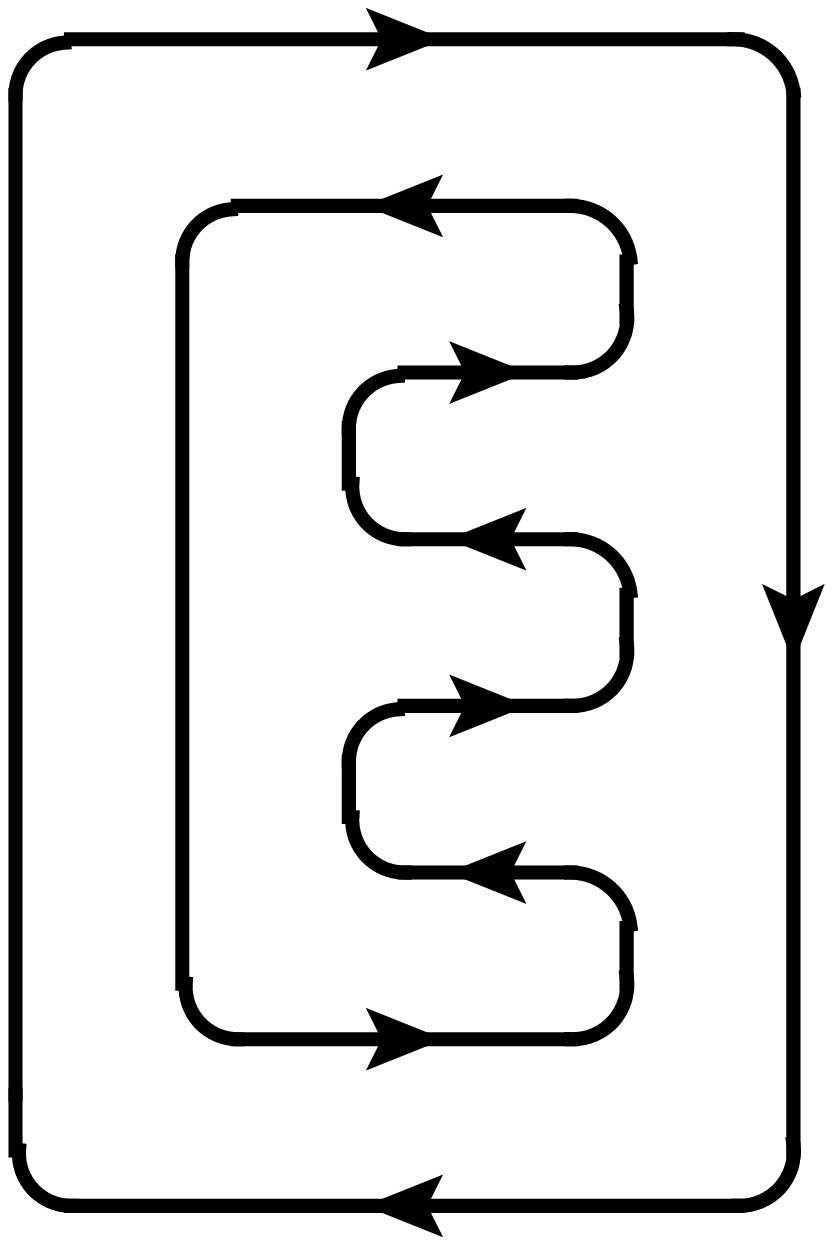}
\hspace{3cm}
\end{minipage}
\begin{minipage}{2cm}
\includegraphics[width = .8\linewidth]{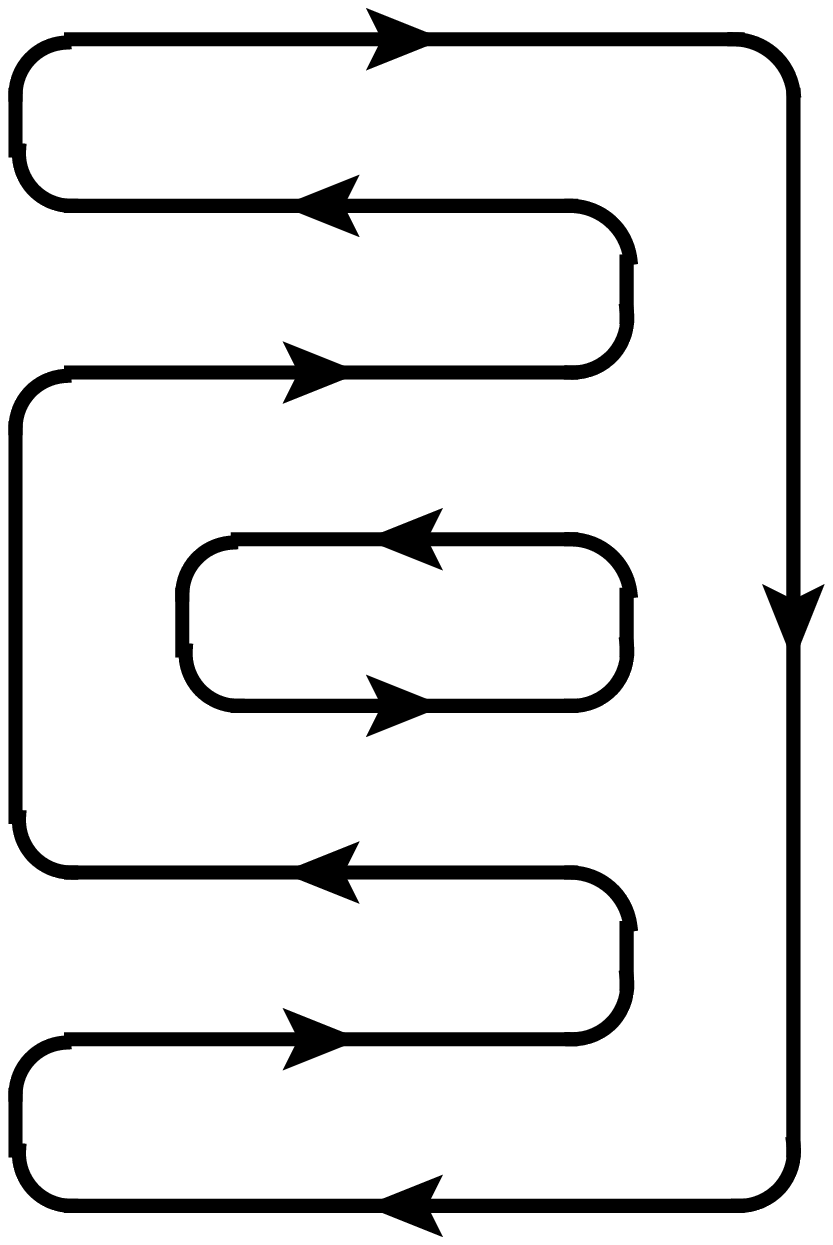}
\hspace{3cm}
\end{minipage}
\begin{minipage}{2cm}
\includegraphics[width = .8\linewidth]{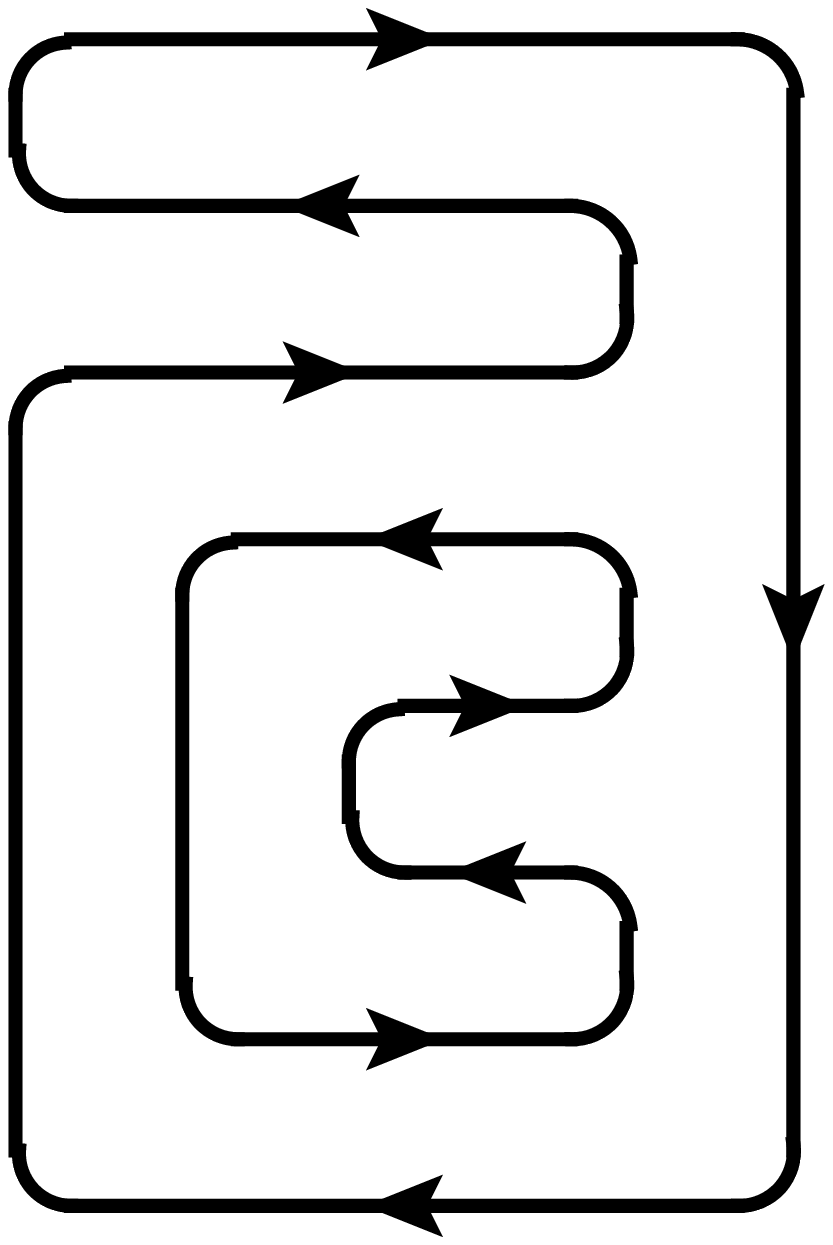}
\hspace{3cm}
\end{minipage}
\begin{minipage}{2cm}
\includegraphics[width = .8\linewidth]{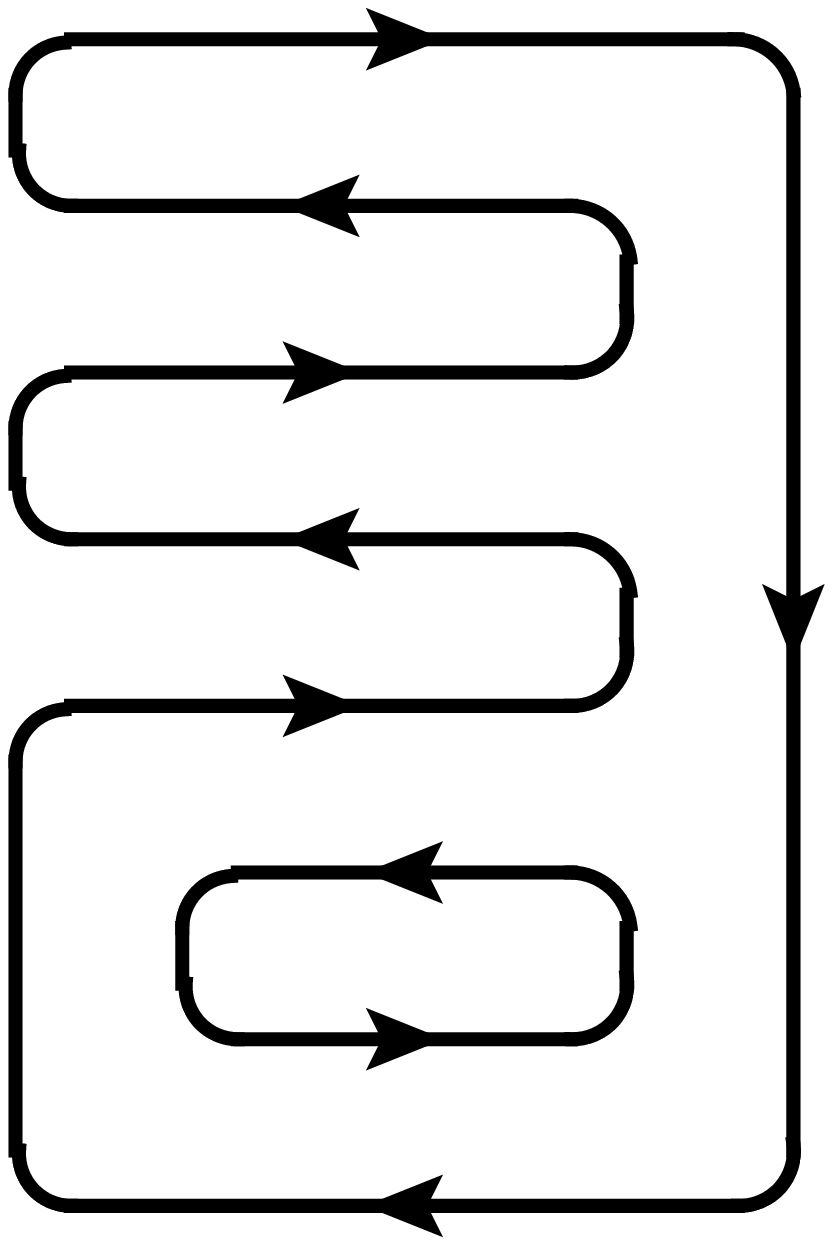}
\hspace{3cm}
\end{minipage}

\begin{minipage}{2cm}
\includegraphics[width = .8\linewidth]{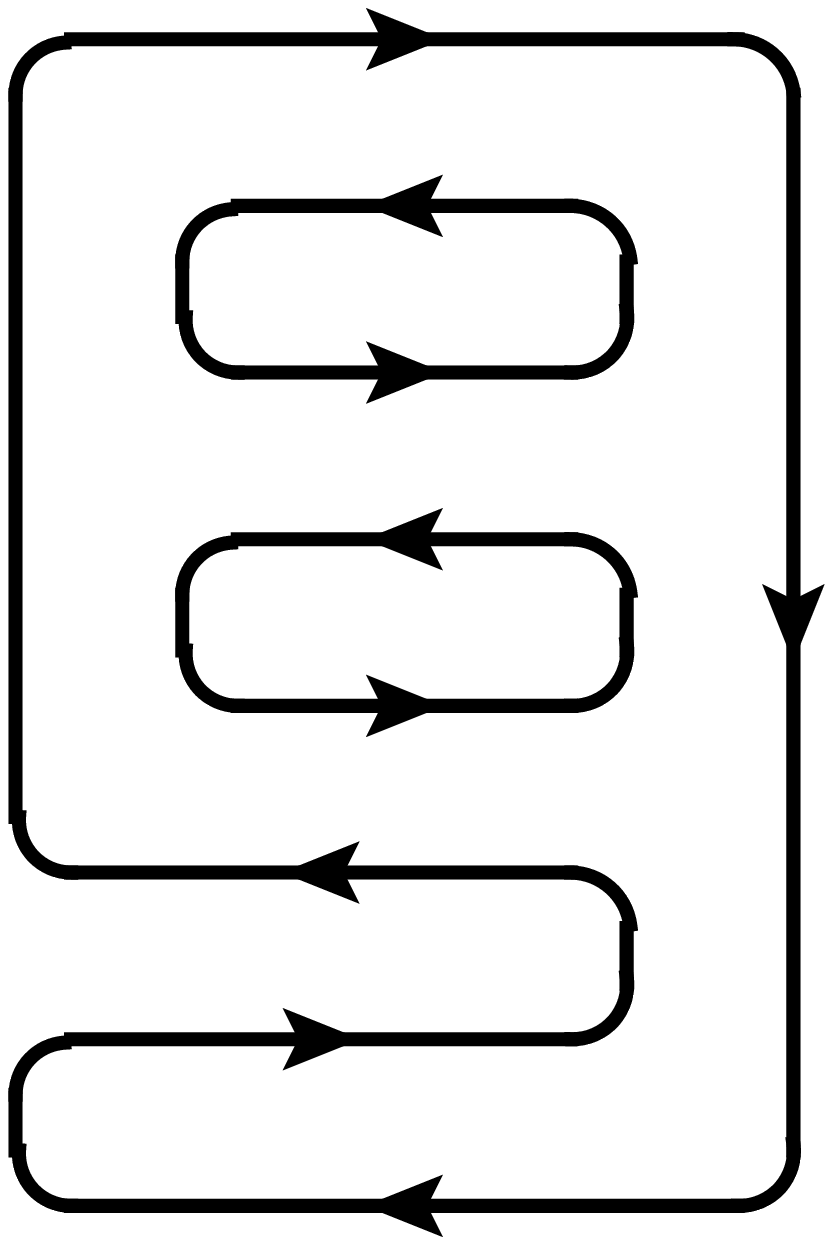}
\hspace{3cm}
\end{minipage}
\begin{minipage}{2cm}
\includegraphics[width = .8\linewidth]{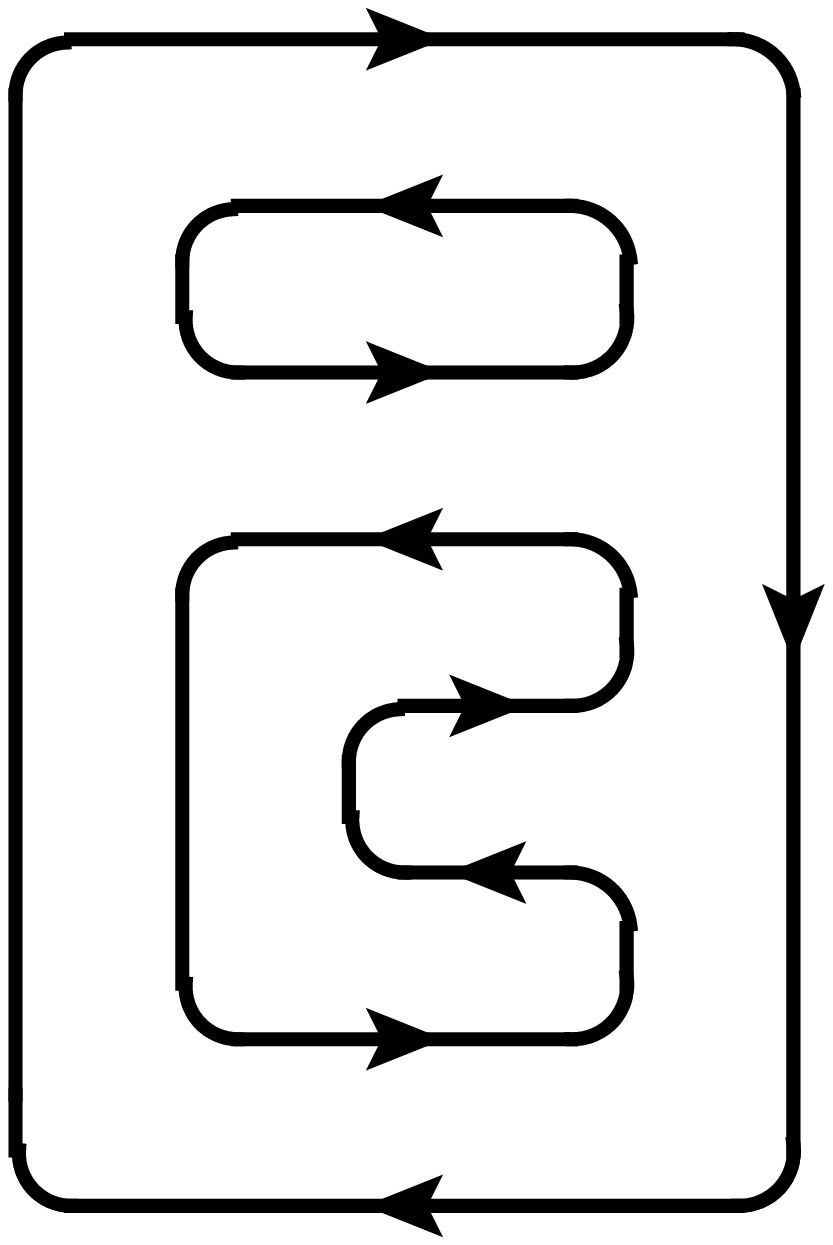}
\hspace{3cm}
\end{minipage}
\begin{minipage}{2cm}
\includegraphics[width = .8\linewidth]{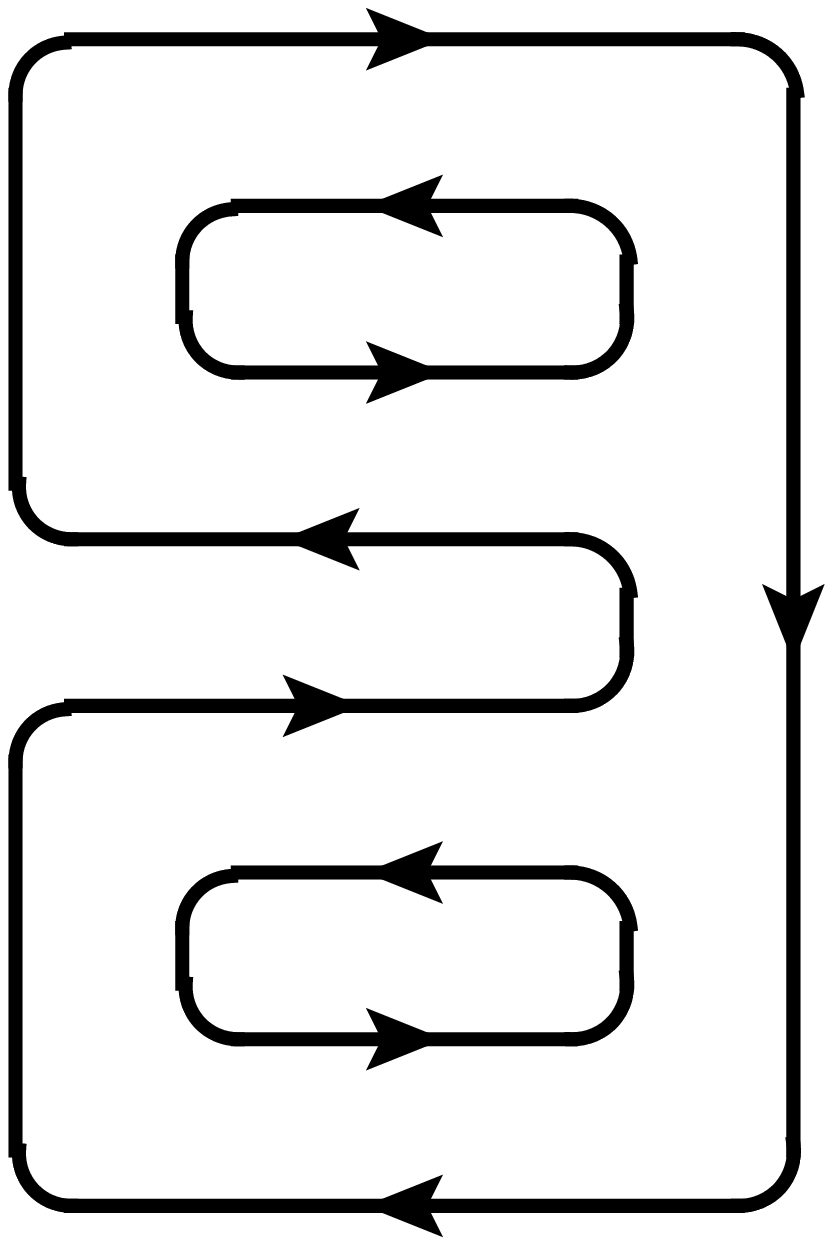}
\hspace{3cm}
\end{minipage}
\begin{minipage}{2cm}
\includegraphics[width = .8\linewidth]{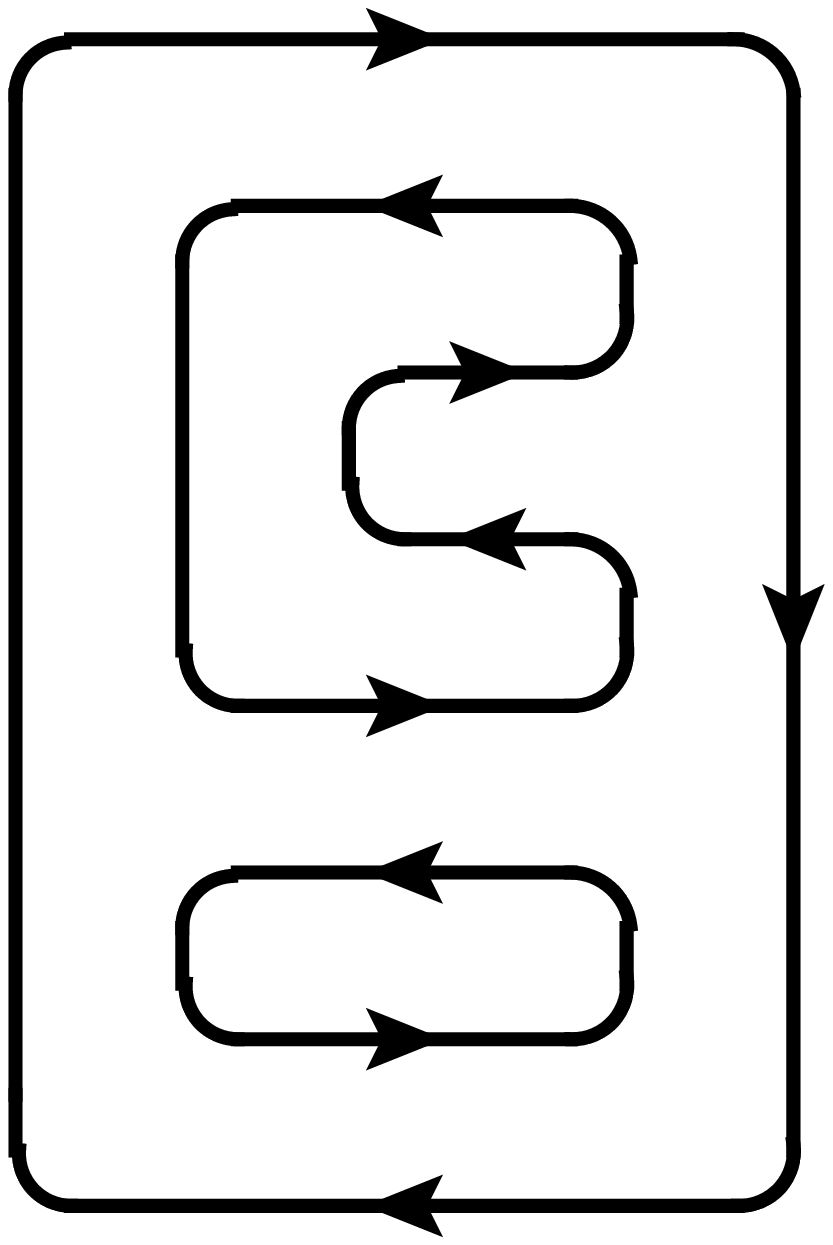}
\hspace{3cm}
\end{minipage}
\begin{minipage}{2cm}
\includegraphics[width = .8\linewidth]{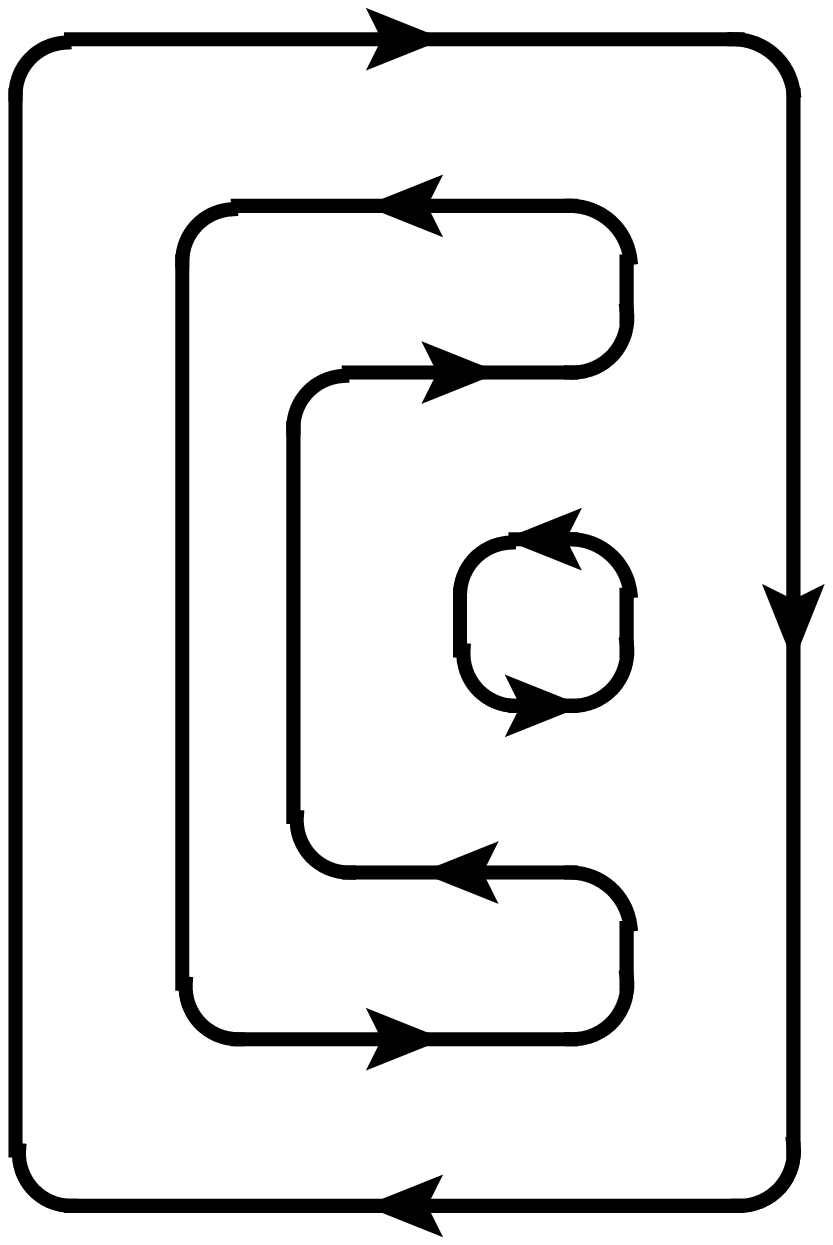}
\hspace{3cm}
\end{minipage}
\begin{minipage}{2cm}
\includegraphics[width = .8\linewidth]{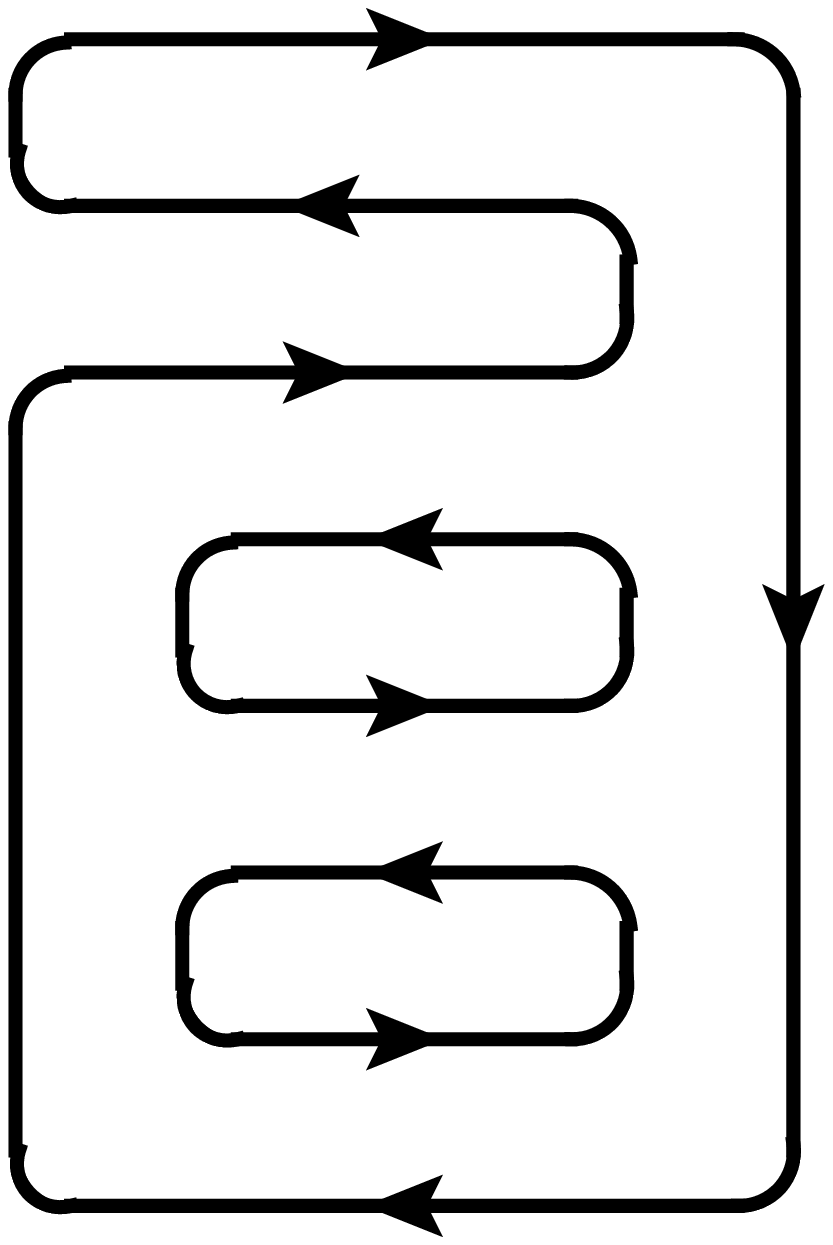}
\hspace{3cm}
\end{minipage}
\begin{minipage}{2cm}
\includegraphics[width = .8\linewidth]{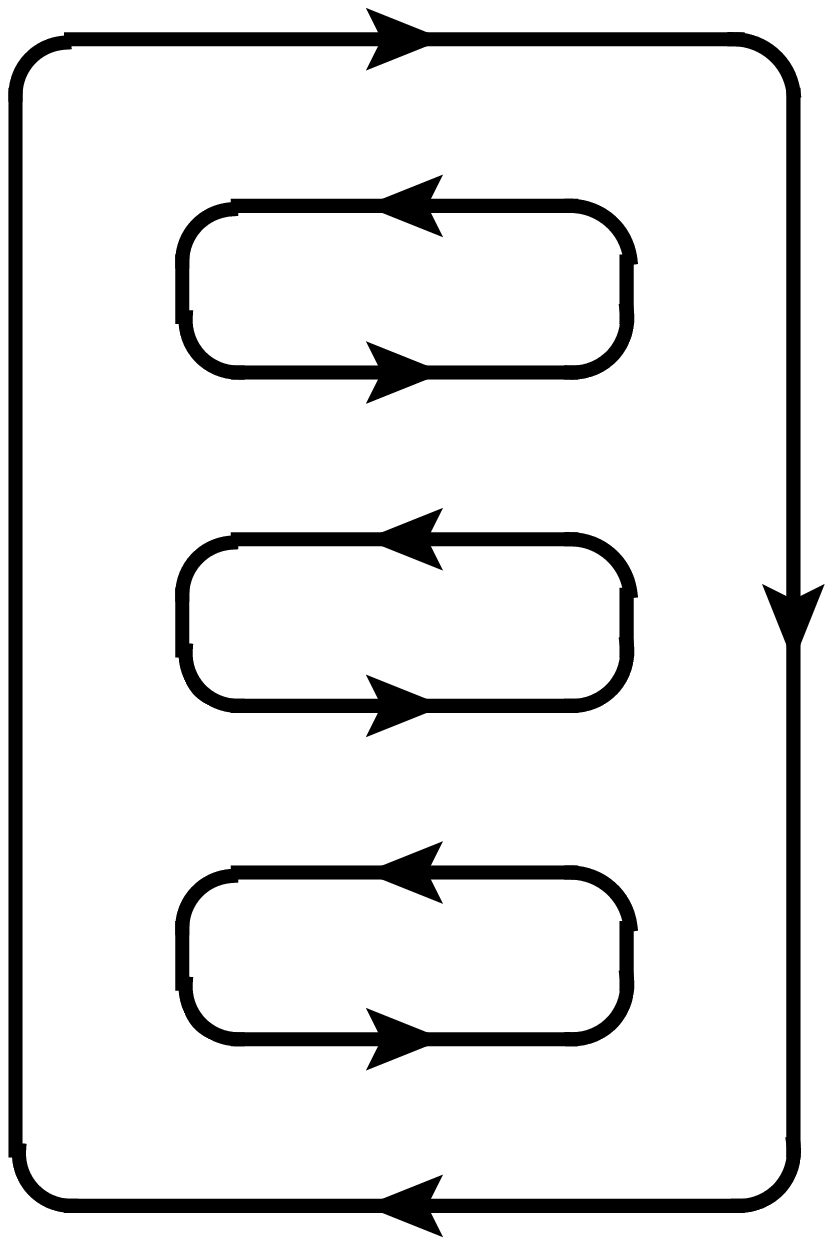}
\hspace{3cm}
\end{minipage}
\caption{The topology of 14 states considered in the evaluation corrections of the octupole.}
\label{topo}
\end{figure}

Based on the formalism developed in Refs. \cite{HiroFujii,Blaizot:2004wv,Dominguez:2012ad}, we calculate the 8-point correlator and conjecture a general expression for 2n-point functions in the MV model. Our calculation is based on the derivation introduced in \cite{Dominguez:2012ad}, in which the result of the 6-point correlator was derived in the large-$N_c$ limit.

The 8-point correlator takes the form

\begin{eqnarray}
&&\frac{1}{N_c}\left\langle\text{Tr}\left[U_{1}U_{2}^{\dagger}U_{3}U_{4}^{\dagger}U_{5}U_{6}^{\dagger}U_{7}U_{8}^{\dagger}\right]\right\rangle\nonumber
\\
&=&\frac{T}{N_c}\sum_{n=0}^\infty\int_{z_1^+<\cdots<z_n^+}\left[N_ca_n(z_1^+,\dots,z_n^+)+N_c^2b_n(z_1^+,\dots,z_n^+)+N_c^3c_n(z_1^+,\dots,z_n^+)+N_c^4d_n(z_1^+,\dots,z_n^+)\right],\label{8}
\end{eqnarray}
where $T$ is the so-called tadpole contribution corresponding to the diagrams where each gluon link attaches to a single Wilson line. The tadpole contribution can be evaluated straightforwardly and gives

\begin{equation}
T=e^{-\frac{C_F}{2}\mu^2\sum\limits_{i=1}^{8}L_{ii}}.
\end{equation}

The rest contributions arising from the states in which every gluon link attaches to two different Wilson lines are presented in the square bracket. Similar to the sextupole calculation, one considers all the possible states contribute in the system of 8 Wilson lines, and finds that there are 24 singlet states in total. In the large-$N_c$ limit \cite{Dominguez:2012ad}, only transitions to states with a higher power of $N_c$, namely, with one more fermion loop, are allowed, which picks out 14 states whose topologies are shown in Fig. \ref{topo}, with $N_c$, $N_c^2$, $N_c^3$ and $N_c^4$ terms representing the first, the next 6, the later 6 and the last diagrams, respectively. Therefore, $b_n$ and $c_n$ in Eq. (\ref{8}) equal the sums of elements in the following column matrices $\bb_{n(6\times 1)}$ and $\bc_{n(6\times 1)}$ respectively, and the corresponding evolution coefficients take the form

\begin{equation}
\left(\begin{matrix}a_n \\ \bb_{n(6\times 1)} \\ \bc_{n(6\times 1)} \\ d_n\end{matrix}\right)=\left[\prod_{i=1}^n\mu^2(z^+_i)\right]M^n\left(\begin{matrix}1\\ \boldsymbol{0}_{(6\times 1)}\\ \boldsymbol{0}_{(6\times 1)}\\ 0\end{matrix}\right),
\end{equation}
in which the transition matrix $M$ can be divided into blocks as

\begin{equation}
M=\left(\begin{matrix}M_1 & 0 & 0&0\\ M_2 &M_3 & 0&0\\ 0 & M_4 & M_5&0\\0&0& M_6 & M_7\end{matrix}\right).\label{block}
\end{equation}

The evolution part of our desired correlator before integrating over the longitudinal coordinate is

\setcounter{MaxMatrixCols}{20}
\begin{equation}
\left(\begin{matrix}
1 & \boldsymbol{N_c}_{(1\times 6)} & \boldsymbol{N_c^2}_{(1\times 6)} & N_c^3
\end{matrix}\right)
M^n
\left(\begin{matrix}
1 \\ \boldsymbol{0}_{(13\times 1)}
\end{matrix}\right),\label{mainpart}
\end{equation}
which picks out the first column of the $n$th power of this matrix

\begin{equation}
\left(\begin{matrix}M_1^n \\ \sum_{i=0}^{n-1}M_3^iM_2M_1^{n-i-1} \\ \sum_{i=0}^{n-2}\sum_{j=0}^{n-i-2}M_5^iM_4M_3^jM_2M_1^{n-i-j-2} \\ \sum_{i=0}^{n-3}\sum_{j=0}^{n-i-3}\sum_{k=0}^{n-i-j-3}M_7^iM_6M_5^jM_4M_3^kM_2M_1^{n-i-j-k-3}\end{matrix}\right).
\end{equation}

As a 14x14 matrix, $M$ has the form
\begin{equation}
M=\begin{pmatrix}
m_{11} & 0 & 0 & 0 & 0 & 0 & 0 & 0 & 0 & 0 & 0 & 0 & 0 & 0\\
m_{21} & m_{22} & 0 & 0 & 0 & 0 & 0 & 0 & 0 & 0 & 0 & 0 & 0 & 0\\
m_{31} & 0 & m_{33} & 0 & 0 & 0 & 0 & 0 & 0 & 0 & 0 & 0 & 0 & 0\\
m_{41} & 0 & 0 & m_{44} & 0 & 0 & 0 & 0 & 0 & 0 & 0 & 0 & 0 & 0\\
m_{51} & 0 & 0 & 0 & m_{55} & 0 & 0 & 0 & 0 & 0 & 0 & 0 & 0 & 0\\
m_{61} & 0 & 0 & 0 & 0 & m_{66} & 0 & 0 & 0 & 0 & 0 & 0 & 0 & 0\\
m_{71} & 0 & 0 & 0 & 0 & 0 & m_{77} & 0 & 0 & 0 & 0 & 0 & 0 & 0\\
0 & m_{82} & m_{83} & 0 & m_{85} & 0 & 0 & m_{88} & 0 & 0 & 0 & 0 & 0 & 0\\
0 & m_{92} & 0 & m_{94} & 0 & m_{96} & 0 & 0 & m_{99} & 0 & 0 & 0 & 0 & 0\\
0 & m_{102} & 0 & 0 & 0 & 0 & m_{107} & 0 & 0 & m_{1010} & 0 & 0 & 0 & 0\\
0 & 0 & m_{113} & m_{114} & 0 & 0 & m_{117} & 0 & 0 & 0 & m_{1111} & 0 & 0 & 0\\
0 & 0 & 0 & m_{124} & m_{125} & 0 & 0 & 0 & 0 & 0 & 0 & m_{1212} & 0 & 0\\
0 & 0 & 0 & 0 & m_{135} & m_{136} & m_{137} & 0 & 0 & 0 & 0 & 0 & m_{1313} & 0\\
0 & 0 & 0 & 0 & 0 & 0 & 0 & m_{148} & m_{149} & m_{1410} & m_{1411} & m_{1412} & m_{1413} & m_{1414}\\
\end{pmatrix},\label{M-matrix}
\end{equation}
which can be written in forms of block matrices in Eq. (\ref{block}) as

\begin{eqnarray}
&&m_{11}=C_FL_{12,34,56,78},
\ m_{1414}=C_FL_{18,23,45,67},
\ M_6=\frac{1}{2}\begin{pmatrix}
F_{1687} & F_{4756} & F_{1485} & F_{2534} & F_{2736} & F_{1283}
\end{pmatrix},
\\
&&M_2=\frac{1}{2}\begin{pmatrix}
F_{1243} \\ F_{1265} \\ F_{1287} \\ F_{3465} \\ F_{3487} \\ F_{5687}
\end{pmatrix},
M_4=\frac{1}{2}\begin{pmatrix}
F_{1465} & F_{2534} & 0 & F_{1263} & 0 & 0
\\
F_{1487} & 0 & F_{2734} & 0 & F_{1283} & 0
\\
F_{5687} & 0 & 0 & 0 & 0 & F_{1243}
\\
0 & F_{1687} & F_{2756} & 0 & 0 & F_{1285}
\\
0 & 0 & F_{3465} & F_{1287} & 0 & 0
\\
0 & 0 & 0 & F_{3687} & F_{4756} & F_{3485}
\end{pmatrix},
\\
&&M_3=C_F\begin{pmatrix}
L_{14,23,56,78} & 0 & 0 & 0 & 0 & 0
\\
0 & L_{16,25,34,78} & 0 & 0 & 0 & 0
\\
0 & 0 & L_{18,27,34,56} & 0 & 0 & 0
\\
0 & 0 & 0 & L_{12,36,45,78} & 0 & 0
\\
0 & 0 & 0 & 0 & L_{12,38,47,56} & 0
\\
0 & 0 & 0 & 0 & 0 & L_{12,34,58,67}
\end{pmatrix},
\\
&&M_5=C_F\begin{pmatrix}
L_{16,23,45,78} & 0 & 0 & 0 & 0 & 0
\\
0 & L_{18,23,47,56} & 0 & 0 & 0 & 0
\\
0 & 0 & L_{14,23,58,67} & 0 & 0 & 0
\\
0 & 0 & 0 & L_{18,25,34,67} & 0 & 0
\\
0 & 0 & 0 & 0 & L_{18,27,36,45} & 0
\\
0 & 0 & 0 & 0 & 0 & L_{12,38,45,67}
\end{pmatrix},
\end{eqnarray}
where

\begin{eqnarray}
L_{ab,cd,ij,kl}=L_{ab}+L_{cd}+L_{ij}+L_{kl}.
\end{eqnarray}

The first column of the $n$th power of $M$ is found to be

\begin{equation}
\begin{pmatrix}
m_{11}^n\\
m_{21}\sum_{i=0}^{n-1}m_{11}^im_{22}^{n-i-1}\\
m_{31}\sum_{i=0}^{n-1}m_{11}^im_{33}^{n-i-1}\\
m_{41}\sum_{i=0}^{n-1}m_{11}^im_{44}^{n-i-1}\\
m_{51}\sum_{i=0}^{n-1}m_{11}^im_{55}^{n-i-1}\\
m_{61}\sum_{i=0}^{n-1}m_{11}^im_{66}^{n-i-1}\\
m_{71}\sum_{i=0}^{n-1}m_{11}^im_{77}^{n-i-1}\\
\sum_{k=2,3,5}m_{8 k}m_{k1}\left[\sum_{i=0}^{n-2}\sum_{j=0}^{n-i-2}m_{11}^im_{kk}^jm_{8 8 }^{n-i-j-2}\right]\\
\sum_{k=2,4,6}m_{9 k}m_{k1}\left[\sum_{i=0}^{n-2}\sum_{j=0}^{n-i-2}m_{11}^im_{kk}^jm_{9 9 }^{n-i-j-2}\right]\\
\sum_{k=2,7  }m_{10k}m_{k1}\left[\sum_{i=0}^{n-2}\sum_{j=0}^{n-i-2}m_{11}^im_{kk}^jm_{1010}^{n-i-j-2}\right]\\
\sum_{k=3,4,7}m_{11k}m_{k1}\left[\sum_{i=0}^{n-2}\sum_{j=0}^{n-i-2}m_{11}^im_{kk}^jm_{1111}^{n-i-j-2}\right]\\
\sum_{k=4,5  }m_{12k}m_{k1}\left[\sum_{i=0}^{n-2}\sum_{j=0}^{n-i-2}m_{11}^im_{kk}^jm_{1212}^{n-i-j-2}\right]\\
\sum_{k=5,6,7}m_{13k}m_{k1}\left[\sum_{i=0}^{n-2}\sum_{j=0}^{n-i-2}m_{11}^im_{kk}^jm_{1313}^{n-i-j-2}\right]\\
\sum_{p=8}^{13}\sum_{q=2}^{7}m_{14p}m_{pq}m_{q1}\left[\sum_{i=0}^{n-3}\sum_{j=0}^{n-i-3}\sum_{k=0}^{n-i-j-3}m_{11}^im_{qq}^jm_{pp}^km_{1414}^{n-i-j-k-3}\right]\\
\end{pmatrix},
\end{equation}
which is derived from the following expression

\begin{equation}
\begin{pmatrix}
m_{11}^n\\
\frac{m_{21}}{m_{11}-m_{22}}\left[m_{11}^n-m_{22}^n\right]\\
\frac{m_{31}}{m_{11}-m_{33}}\left[m_{11}^n-m_{33}^n\right]\\
\frac{m_{41}}{m_{11}-m_{44}}\left[m_{11}^n-m_{44}^n\right]\\
\frac{m_{51}}{m_{11}-m_{55}}\left[m_{11}^n-m_{55}^n\right]\\
\frac{m_{61}}{m_{11}-m_{66}}\left[m_{11}^n-m_{66}^n\right]\\
\frac{m_{71}}{m_{11}-m_{77}}\left[m_{11}^n-m_{77}^n\right]\\
\sum_{k=2,3,5}m_{8k}m_{k1}\left[\frac{m_{11}^n}{(m_{11}-m_{kk})(m_{11}-m_{88})}+\frac{m_{kk}^n}{(m_{kk}-m_{11})(m_{kk}-m_{88})}+\frac{m_{88}^n}{(m_{88}-m_{kk})(m_{88}-m_{11})}\right]\\
\sum_{k=2,4,6}m_{9k}m_{k1}\left[\frac{m_{11}^n}{(m_{11}-m_{kk})(m_{11}-m_{99})}+\frac{m_{kk}^n}{(m_{kk}-m_{11})(m_{kk}-m_{99})}+\frac{m_{99}^n}{(m_{99}-m_{kk})(m_{99}-m_{11})}\right]\\
\sum_{k=2,7}m_{10k}m_{k1}\left[\frac{m_{11}^n}{(m_{11}-m_{kk})(m_{11}-m_{1010})}+\frac{m_{kk}^n}{(m_{kk}-m_{11})(m_{kk}-m_{1010})}+\frac{m_{1010}^n}{(m_{1010}-m_{kk})(m_{1010}-m_{11})}\right]\\
\sum_{k=3,4,7}m_{11k}m_{k1}\left[\frac{m_{11}^n}{(m_{11}-m_{kk})(m_{11}-m_{1111})}+\frac{m_{kk}^n}{(m_{kk}-m_{11})(m_{kk}-m_{1111})}+\frac{m_{1111}^n}{(m_{1111}-m_{kk})(m_{1111}-m_{11})}\right]\\
\sum_{k=4,5}m_{12k}m_{k1}\left[\frac{m_{11}^n}{(m_{11}-m_{kk})(m_{11}-m_{1212})}+\frac{m_{kk}^n}{(m_{kk}-m_{11})(m_{kk}-m_{1212})}+\frac{m_{1212}^n}{(m_{1212}-m_{kk})(m_{1212}-m_{11})}\right]\\
\sum_{k=5,6,7}m_{13k}m_{k1}\left[\frac{m_{11}^n}{(m_{11}-m_{kk})(m_{11}-m_{1313})}+\frac{m_{kk}^n}{(m_{kk}-m_{11})(m_{kk}-m_{1313})}+\frac{m_{1313}^n}{(m_{1313}-m_{kk})(m_{1313}-m_{11})}\right]\\
\sum_{p=8}^{13}\sum_{q=2}^{7}m_{14p}m_{pq}m_{q1}[\frac{m_{11}^n}{(m_{11}-m_{qq})(m_{11}-m_{pp})(m_{11}-m_{1414})}+\frac{m_{qq}^n}{(m_{qq}-m_{11})(m_{qq}-m_{pp})(m_{qq}-m_{1414})}\\
+\frac{m_{pp}^n}{(m_{pp}-m_{11})(m_{pp}-m_{qq})(m_{pp}-m_{1414})}+\frac{m_{1414}^n}{(m_{1414}-m_{11})(m_{1414}-m_{qq})(m_{1414}-m_{pp})}]\\
\end{pmatrix}.\label{nth1stcol}
\end{equation}

Plugging the first column into the above evolution part in Eq. (\ref{mainpart}), integrating over the longitudinal coordinate, summing over $n$ with a factor of $\frac{1}{n!}$, and including the tadpole contribution, one obtains

\begin{eqnarray}
&&\frac{1}{N_c}\left\langle\text{Tr}\left[U_{1}U_{2}^{\dagger}U_{3}U_{4}^{\dagger}U_{5}U_{6}^{\dagger}U_{7}U_{8}^{\dagger}\right]\right\rangle\nonumber
\\
&=&e^{-\Gamma_{12}-\Gamma_{34}-\Gamma_{56}-\Gamma_{78}}
-\frac{F_{1234}}{F_{1324}}\left[e^{-\Gamma_{12}-\Gamma_{34}}-e^{-\Gamma_{14}-\Gamma_{32}}\right]e^{-\Gamma_{56}-\Gamma_{78}}
-\frac{F_{1256}}{F_{1526}}\left[e^{-\Gamma_{12}-\Gamma_{56}}-e^{-\Gamma_{16}-\Gamma_{52}}\right]e^{-\Gamma_{34}-\Gamma_{78}}\nonumber
\\
&&-\frac{F_{1278}}{F_{1728}}\left[e^{-\Gamma_{12}-\Gamma_{78}}-e^{-\Gamma_{18}-\Gamma_{72}}\right]e^{-\Gamma_{34}-\Gamma_{56}}
-\frac{F_{3456}}{F_{3546}}\left[e^{-\Gamma_{34}-\Gamma_{56}}-e^{-\Gamma_{36}-\Gamma_{54}}\right]e^{-\Gamma_{12}-\Gamma_{78}}\nonumber
\\
&&-\frac{F_{3478}}{F_{3748}}\left[e^{-\Gamma_{34}-\Gamma_{78}}-e^{-\Gamma_{38}-\Gamma_{74}}\right]e^{-\Gamma_{12}-\Gamma_{56}}
-\frac{F_{5678}}{F_{5768}}\left[e^{-\Gamma_{56}-\Gamma_{78}}-e^{-\Gamma_{58}-\Gamma_{76}}\right]e^{-\Gamma_{12}-\Gamma_{34}}\nonumber
\\
&&+F_{1243}F_{1465}\left[\frac{e^{-\Gamma_{12}-\Gamma_{34}-\Gamma_{56}}}{F_{1324}G_1}-\frac{e^{-\Gamma_{14}-\Gamma_{32}-\Gamma_{56}}}{F_{1324}F_{1546}}
+\frac{e^{-\Gamma_{16}-\Gamma_{32}-\Gamma_{54}}}{F_{1546}G_1}\right]e^{-\Gamma_{78}}\nonumber
\\
&&+F_{1265}F_{2534}\left[\frac{e^{-\Gamma_{12}-\Gamma_{34}-\Gamma_{56}}}{F_{1526}G_1}-\frac{e^{-\Gamma_{16}-\Gamma_{34}-\Gamma_{52}}}{F_{1526}F_{2453}}
+\frac{e^{-\Gamma_{16}-\Gamma_{32}-\Gamma_{54}}}{F_{2453}G_1}\right]e^{-\Gamma_{78}}\nonumber
\\
&&+F_{3465}F_{1263}\left[\frac{e^{-\Gamma_{12}-\Gamma_{34}-\Gamma_{56}}}{F_{3546}G_1}-\frac{e^{-\Gamma_{12}-\Gamma_{36}-\Gamma_{54}}}{F_{3546}F_{1326}}
+\frac{e^{-\Gamma_{16}-\Gamma_{32}-\Gamma_{54}}}{F_{1326}G_1}\right]e^{-\Gamma_{78}}\nonumber
\\
&&+F_{1243}F_{1487}\left[\frac{e^{-\Gamma_{12}-\Gamma_{34}-\Gamma_{78}}}{F_{1324}G_2}-\frac{e^{-\Gamma_{14}-\Gamma_{32}-\Gamma_{78}}}{F_{1324}F_{1748}}
+\frac{e^{-\Gamma_{18}-\Gamma_{32}-\Gamma_{74}}}{F_{1748}G_2}\right]e^{-\Gamma_{56}}\nonumber
\\
&&+F_{1287}F_{2734}\left[\frac{e^{-\Gamma_{12}-\Gamma_{34}-\Gamma_{78}}}{F_{1728}G_2}-\frac{e^{-\Gamma_{18}-\Gamma_{72}-\Gamma_{34}}}{F_{1728}F_{2473}}
+\frac{e^{-\Gamma_{18}-\Gamma_{32}-\Gamma_{74}}}{F_{2473}G_2}\right]e^{-\Gamma_{56}}\nonumber
\\
&&+F_{3487}F_{1283}\left[\frac{e^{-\Gamma_{12}-\Gamma_{34}-\Gamma_{78}}}{F_{3748}G_2}-\frac{e^{-\Gamma_{12}-\Gamma_{38}-\Gamma_{74}}}{F_{3748}F_{1328}}
+\frac{e^{-\Gamma_{18}-\Gamma_{32}-\Gamma_{74}}}{F_{1328}G_2}\right]e^{-\Gamma_{56}}\nonumber
\\
&&+F_{1243}F_{5687}\left[
\frac{e^{-\Gamma_{12}-\Gamma_{34}-\Gamma_{56}-\Gamma_{78}}}{F_{1324}G_3}-\frac{e^{-\Gamma_{14}-\Gamma_{32}-\Gamma_{56}-\Gamma_{78}}}{F_{1324}F_{5768}}
+\frac{e^{-\Gamma_{14}-\Gamma_{32}-\Gamma_{58}-\Gamma_{76}}}{F_{5768}G_3}\right]\nonumber
\\
&&+F_{5687}F_{1243}\left[
\frac{e^{-\Gamma_{12}-\Gamma_{34}-\Gamma_{56}-\Gamma_{78}}}{F_{5768}G_3}-\frac{e^{-\Gamma_{12}-\Gamma_{34}-\Gamma_{58}-\Gamma_{76}}}{F_{5768}F_{1324}}
+\frac{e^{-\Gamma_{14}-\Gamma_{32}-\Gamma_{58}-\Gamma_{76}}}{F_{1324}G_3}\right]\nonumber
\\
&&+F_{1265}F_{1687}\left[
\frac{e^{-\Gamma_{12}-\Gamma_{56}-\Gamma_{78}}}{F_{1526}G_4}-\frac{e^{-\Gamma_{16}-\Gamma_{52}-\Gamma_{78}}}{F_{1526}F_{1768}}
+\frac{e^{-\Gamma_{18}-\Gamma_{52}-\Gamma_{76}}}{F_{1768}G_4}\right]e^{-\Gamma_{34}}\nonumber
\\
&&+F_{1287}F_{2756}\left[
\frac{e^{-\Gamma_{12}-\Gamma_{56}-\Gamma_{78}}}{F_{1728}G_4}-\frac{e^{-\Gamma_{18}-\Gamma_{72}-\Gamma_{56}}}{F_{1728}F_{2675}}
+\frac{e^{-\Gamma_{18}-\Gamma_{52}-\Gamma_{76}}}{F_{2675}G_4}\right]e^{-\Gamma_{34}}\nonumber
\\
&&+F_{5687}F_{1285}\left[
\frac{e^{-\Gamma_{12}-\Gamma_{56}-\Gamma_{78}}}{F_{5768}G_4}-\frac{e^{-\Gamma_{12}-\Gamma_{58}-\Gamma_{76}}}{F_{5768}F_{1528}}
+\frac{e^{-\Gamma_{18}-\Gamma_{52}-\Gamma_{76}}}{F_{1528}G_4}\right]e^{-\Gamma_{34}}\nonumber
\\
&&+F_{1287}F_{3465}\left[
\frac{e^{-\Gamma_{12}-\Gamma_{34}-\Gamma_{56}-\Gamma_{78}}}{F_{1728}G_5}-\frac{e^{-\Gamma_{18}-\Gamma_{72}-\Gamma_{34}-\Gamma_{56}}}{F_{1728}F_{3546}}
+\frac{e^{-\Gamma_{18}-\Gamma_{72}-\Gamma_{36}-\Gamma_{54}}}{F_{3546}G_5}\right]\nonumber
\\
&&+F_{3465}F_{1287}\left[
\frac{e^{-\Gamma_{12}-\Gamma_{34}-\Gamma_{56}-\Gamma_{78}}}{F_{3546}G_5}-\frac{e^{-\Gamma_{12}-\Gamma_{36}-\Gamma_{54}-\Gamma_{78}}}{F_{3546}F_{1728}}
+\frac{e^{-\Gamma_{18}-\Gamma_{72}-\Gamma_{36}-\Gamma_{54}}}{F_{1728}G_5}\right]\nonumber
\\
&&+F_{3465}F_{3687}\left[
\frac{e^{-\Gamma_{34}-\Gamma_{56}-\Gamma_{78}}}{F_{3546}G_6}-\frac{e^{-\Gamma_{36}-\Gamma_{54}-\Gamma_{78}}}{F_{3546}F_{3768}}
+\frac{e^{-\Gamma_{38}-\Gamma_{54}-\Gamma_{76}}}{F_{3768}G_6}\right]e^{-\Gamma_{12}}\nonumber
\\
&&+F_{3487}F_{4756}\left[
\frac{e^{-\Gamma_{34}-\Gamma_{56}-\Gamma_{78}}}{F_{3748}G_6}-\frac{e^{-\Gamma_{38}-\Gamma_{74}-\Gamma_{56}}}{F_{3748}F_{4675}}
+\frac{e^{-\Gamma_{38}-\Gamma_{54}-\Gamma_{76}}}{F_{4675}G_6}\right]e^{-\Gamma_{12}}\nonumber
\\
&&+F_{5687}F_{3485}\left[
\frac{e^{-\Gamma_{34}-\Gamma_{56}-\Gamma_{78}}}{F_{5768}G_6}-\frac{e^{-\Gamma_{34}-\Gamma_{58}-\Gamma_{76}}}{F_{5768}F_{3548}}
+\frac{e^{-\Gamma_{38}-\Gamma_{54}-\Gamma_{76}}}{F_{3548}G_6}\right]e^{-\Gamma_{12}}\nonumber
\\
&&+F_{1243}F_{1465}F_{1687}\left[
\frac{e^{-\Gamma_{12}-\Gamma_{34}-\Gamma_{56}-\Gamma_{78}}}{F_{1324}G_1K}-\frac{e^{-\Gamma_{14}-\Gamma_{32}-\Gamma_{56}-\Gamma_{78}}}{F_{1324}F_{1546}H_1}+
\frac{e^{-\Gamma_{16}-\Gamma_{32}-\Gamma_{54}-\Gamma_{78}}}{G_1F_{1546}F_{1768}}-\frac{e^{-\Gamma_{18}-\Gamma_{32}-\Gamma_{54}-\Gamma_{76}}}{KH_1F_{1768}}\right]\nonumber \\
&&+F_{1243}F_{1487}F_{4756}\left[
\frac{e^{-\Gamma_{12}-\Gamma_{34}-\Gamma_{56}-\Gamma_{78}}}{F_{1324}G_2K}-\frac{e^{-\Gamma_{14}-\Gamma_{32}-\Gamma_{56}-\Gamma_{78}}}{F_{1324}F_{1748}H_1}+
\frac{e^{-\Gamma_{18}-\Gamma_{32}-\Gamma_{74}-\Gamma_{56}}}{G_2F_{1748}F_{4675}}-\frac{e^{-\Gamma_{18}-\Gamma_{32}-\Gamma_{54}-\Gamma_{76}}}{KH_1F_{4675}}\right]\nonumber \\
&&+F_{1243}F_{5687}F_{1485}\left[
\frac{e^{-\Gamma_{12}-\Gamma_{34}-\Gamma_{56}-\Gamma_{78}}}{F_{1324}G_3K}-\frac{e^{-\Gamma_{14}-\Gamma_{32}-\Gamma_{56}-\Gamma_{78}}}{F_{1324}F_{5768}H_1}+
\frac{e^{-\Gamma_{14}-\Gamma_{32}-\Gamma_{58}-\Gamma_{76}}}{G_3F_{5768}F_{1548}}-\frac{e^{-\Gamma_{18}-\Gamma_{32}-\Gamma_{54}-\Gamma_{76}}}{KH_1F_{1548}}\right]\nonumber \\
&&+F_{1265}F_{2534}F_{1687}\left[
\frac{e^{-\Gamma_{12}-\Gamma_{34}-\Gamma_{56}-\Gamma_{78}}}{F_{1526}G_1K}-\frac{e^{-\Gamma_{16}-\Gamma_{52}-\Gamma_{34}-\Gamma_{78}}}{F_{1526}F_{2453}H_2}+
\frac{e^{-\Gamma_{16}-\Gamma_{32}-\Gamma_{54}-\Gamma_{78}}}{G_1F_{2453}F_{1768}}-\frac{e^{-\Gamma_{18}-\Gamma_{32}-\Gamma_{54}-\Gamma_{76}}}{KH_2F_{1768}}\right]\nonumber \\
&&+F_{1265}F_{1687}F_{2534}\left[
\frac{e^{-\Gamma_{12}-\Gamma_{34}-\Gamma_{56}-\Gamma_{78}}}{F_{1526}G_4K}-\frac{e^{-\Gamma_{16}-\Gamma_{52}-\Gamma_{34}-\Gamma_{78}}}{F_{1526}F_{1768}H_2}+
\frac{e^{-\Gamma_{18}-\Gamma_{52}-\Gamma_{34}-\Gamma_{76}}}{G_4F_{1768}F_{2453}}-\frac{e^{-\Gamma_{18}-\Gamma_{32}-\Gamma_{54}-\Gamma_{76}}}{KH_2F_{2453}}\right]\nonumber \\
&&+F_{1287}F_{2734}F_{4756}\left[
\frac{e^{-\Gamma_{12}-\Gamma_{34}-\Gamma_{56}-\Gamma_{78}}}{F_{1728}G_2K}-\frac{e^{-\Gamma_{18}-\Gamma_{72}-\Gamma_{34}-\Gamma_{56}}}{F_{1728}F_{2473}H_3}+
\frac{e^{-\Gamma_{18}-\Gamma_{32}-\Gamma_{54}-\Gamma_{56}}}{G_2F_{2473}F_{4675}}-\frac{e^{-\Gamma_{18}-\Gamma_{32}-\Gamma_{54}-\Gamma_{76}}}{KH_3F_{4675}}\right]\nonumber \\
&&+F_{1287}F_{2756}F_{2534}\left[
\frac{e^{-\Gamma_{12}-\Gamma_{34}-\Gamma_{56}-\Gamma_{78}}}{F_{1728}G_4K}-\frac{e^{-\Gamma_{18}-\Gamma_{72}-\Gamma_{34}-\Gamma_{56}}}{F_{1728}F_{2675}H_3}+
\frac{e^{-\Gamma_{18}-\Gamma_{52}-\Gamma_{34}-\Gamma_{76}}}{G_4F_{2675}F_{2453}}-\frac{e^{-\Gamma_{18}-\Gamma_{32}-\Gamma_{54}-\Gamma_{76}}}{KH_3F_{2453}}\right]\nonumber \\
&&+F_{1287}F_{3465}F_{2736}\left[
\frac{e^{-\Gamma_{12}-\Gamma_{34}-\Gamma_{56}-\Gamma_{78}}}{F_{1728}G_5K}-\frac{e^{-\Gamma_{18}-\Gamma_{72}-\Gamma_{34}-\Gamma_{56}}}{F_{1728}F_{3546}H_3}+
\frac{e^{-\Gamma_{18}-\Gamma_{72}-\Gamma_{36}-\Gamma_{54}}}{G_5F_{3546}F_{2673}}-\frac{e^{-\Gamma_{18}-\Gamma_{32}-\Gamma_{54}-\Gamma_{76}}}{KH_3F_{2673}}\right]\nonumber \\
&&+F_{3465}F_{1263}F_{1687}\left[
\frac{e^{-\Gamma_{12}-\Gamma_{34}-\Gamma_{56}-\Gamma_{78}}}{F_{3546}G_1K}-\frac{e^{-\Gamma_{12}-\Gamma_{36}-\Gamma_{54}-\Gamma_{78}}}{F_{3546}F_{1326}H_4}+
\frac{e^{-\Gamma_{16}-\Gamma_{32}-\Gamma_{54}-\Gamma_{78}}}{G_1F_{1326}F_{1768}}-\frac{e^{-\Gamma_{18}-\Gamma_{32}-\Gamma_{54}-\Gamma_{76}}}{KH_4F_{1768}}\right]\nonumber \\
&&+F_{3465}F_{1287}F_{2736}\left[
\frac{e^{-\Gamma_{12}-\Gamma_{34}-\Gamma_{56}-\Gamma_{78}}}{F_{3546}G_5K}-\frac{e^{-\Gamma_{12}-\Gamma_{36}-\Gamma_{54}-\Gamma_{78}}}{F_{3546}F_{1728}H_4}+
\frac{e^{-\Gamma_{18}-\Gamma_{72}-\Gamma_{36}-\Gamma_{54}}}{G_5F_{1728}F_{2673}}-\frac{e^{-\Gamma_{18}-\Gamma_{32}-\Gamma_{54}-\Gamma_{76}}}{KH_4F_{2673}}\right]\nonumber \\
&&+F_{3465}F_{3687}F_{1283}\left[
\frac{e^{-\Gamma_{12}-\Gamma_{34}-\Gamma_{56}-\Gamma_{78}}}{F_{3546}G_6K}-\frac{e^{-\Gamma_{12}-\Gamma_{36}-\Gamma_{54}-\Gamma_{78}}}{F_{3546}F_{3768}H_4}+
\frac{e^{-\Gamma_{12}-\Gamma_{38}-\Gamma_{54}-\Gamma_{76}}}{G_6F_{3768}F_{1328}}-\frac{e^{-\Gamma_{18}-\Gamma_{32}-\Gamma_{54}-\Gamma_{76}}}{KH_4F_{1328}}\right]\nonumber \\
&&+F_{3487}F_{1283}F_{4756}\left[
\frac{e^{-\Gamma_{12}-\Gamma_{34}-\Gamma_{56}-\Gamma_{78}}}{F_{3748}G_2K}-\frac{e^{-\Gamma_{12}-\Gamma_{38}-\Gamma_{74}-\Gamma_{56}}}{F_{3748}F_{1328}H_5}+
\frac{e^{-\Gamma_{18}-\Gamma_{32}-\Gamma_{74}-\Gamma_{56}}}{G_2F_{1328}F_{4675}}-\frac{e^{-\Gamma_{18}-\Gamma_{32}-\Gamma_{54}-\Gamma_{76}}}{KH_5F_{4675}}\right]\nonumber \\
&&+F_{3487}F_{4756}F_{1283}\left[
\frac{e^{-\Gamma_{12}-\Gamma_{34}-\Gamma_{56}-\Gamma_{78}}}{F_{3748}G_6K}-\frac{e^{-\Gamma_{12}-\Gamma_{38}-\Gamma_{74}-\Gamma_{56}}}{F_{3748}F_{4675}H_5}+
\frac{e^{-\Gamma_{12}-\Gamma_{38}-\Gamma_{54}-\Gamma_{76}}}{G_6F_{4675}F_{1328}}-\frac{e^{-\Gamma_{18}-\Gamma_{32}-\Gamma_{54}-\Gamma_{76}}}{KH_5F_{1328}}\right]\nonumber \\
&&+F_{5687}F_{1243}F_{1485}\left[
\frac{e^{-\Gamma_{12}-\Gamma_{34}-\Gamma_{56}-\Gamma_{78}}}{F_{5768}G_3K}-\frac{e^{-\Gamma_{12}-\Gamma_{34}-\Gamma_{58}-\Gamma_{76}}}{F_{5768}F_{1324}H_6}+
\frac{e^{-\Gamma_{14}-\Gamma_{32}-\Gamma_{58}-\Gamma_{76}}}{G_3F_{1324}F_{1548}}-\frac{e^{-\Gamma_{18}-\Gamma_{32}-\Gamma_{54}-\Gamma_{76}}}{KH_6F_{1548}}\right]\nonumber \\
&&+F_{5687}F_{1285}F_{2534}\left[
\frac{e^{-\Gamma_{12}-\Gamma_{34}-\Gamma_{56}-\Gamma_{78}}}{F_{5768}G_4K}-\frac{e^{-\Gamma_{12}-\Gamma_{34}-\Gamma_{58}-\Gamma_{76}}}{F_{5768}F_{1528}H_6}+
\frac{e^{-\Gamma_{18}-\Gamma_{25}-\Gamma_{43}-\Gamma_{76}}}{G_4F_{1528}F_{2453}}-\frac{e^{-\Gamma_{18}-\Gamma_{32}-\Gamma_{54}-\Gamma_{76}}}{KH_6F_{2453}}\right]\nonumber \\
&&+F_{5687}F_{3485}F_{1283}\left[
\frac{e^{-\Gamma_{12}-\Gamma_{34}-\Gamma_{56}-\Gamma_{78}}}{F_{5768}G_6K}-\frac{e^{-\Gamma_{12}-\Gamma_{34}-\Gamma_{58}-\Gamma_{76}}}{F_{5768}F_{3548}H_6}+
\frac{e^{-\Gamma_{12}-\Gamma_{38}-\Gamma_{54}-\Gamma_{76}}}{G_6F_{3548}F_{1328}}-\frac{e^{-\Gamma_{18}-\Gamma_{32}-\Gamma_{54}-\Gamma_{76}}}{KH_6F_{1328}}\right],
\label{8result}
\end{eqnarray}
where

\begin{eqnarray}
&&G_1=L_{12}+L_{34}+L_{56}-L_{16}-L_{32}-L_{54},
\\
&&G_2=L_{12}+L_{34}+L_{78}-L_{18}-L_{32}-L_{74},
\\
&&G_3=L_{12}+L_{34}+L_{56}+L_{78}-L_{14}-L_{32}-L_{58}-L_{76},
\\
&&G_4=L_{12}+L_{56}+L_{78}-L_{18}-L_{52}-L_{76},
\\
&&G_5=L_{12}+L_{34}+L_{56}+L_{78}-L_{18}-L_{72}-L_{36}-L_{54},
\\
&&G_6=L_{34}+L_{56}+L_{78}-L_{38}-L_{54}-L_{76},
\\
&&H_1=L_{14}+L_{56}+L_{78}-L_{18}-L_{54}-L_{76},
\\
&&H_2=L_{16}+L_{52}+L_{34}+L_{78}-L_{18}-L_{32}-L_{54}-L_{76},
\\
&&H_3=L_{72}+L_{34}+L_{56}-L_{32}-L_{54}-L_{76},
\\
&&H_4=L_{12}+L_{36}+L_{78}-L_{18}-L_{32}-L_{76},
\\
&&H_5=L_{12}+L_{38}+L_{74}+L_{56}-L_{18}-L_{32}-L_{54}-L_{76},
\\
&&H_6=L_{12}+L_{34}+L_{58}-L_{18}-L_{32}-L_{54},
\\
&&K=L_{12}+L_{34}+L_{56}+L_{78}-L_{18}-L_{32}-L_{54}-L_{76}.
\end{eqnarray}

Along with 2, 4 and 6-point correlators described above and according to patterns that we observe, we can conjecture a general expression of the 2n-point correlator given as

\begin{eqnarray}
&&\frac{1}{N_c}\left\langle\text{Tr}\left[U_{1}U_{2}^{\dagger}...U_{2n-1}U_{2n}^{\dagger}\right]\right\rangle\nonumber
\\
&=&\left[e^{C_F\mu^2A_1} +\sum_{k=2}^n   \sum_{I_k}
        \left[\prod_{i=1}^{k-1}F_{a_ib_ic_id_i}      \sum_{l=1}^k\frac{e^{C_F\mu^2A_l}}{\prod\limits_{j=1;j\neq l}^k(A_l-A_j)}
        \right]
   \right]
e^{-\frac{C_F}{2}\mu^2\sum\limits_{r=1}^{2n}L_{rr}},\label{2n-point}
\end{eqnarray}
where $I_k$ represents summing over all possible permutations $a_1,b_1,c_1,d_1,a_2,b_2,c_2,d_2,...,a_{k-1},b_{k-1},c_{k-1},d_{k-1}$,
$A_1,A_2,...,A_k$ that satisfy the following conditions

\begin{eqnarray}
&&a_i<d_i; \ (a_i,b_i),(d_i,c_i)\in U_i,
\\
&&A_i=\sum_{(a,b)\in U_i}L_{ab},
\end{eqnarray}
where $U_i$ satisfies

\begin{eqnarray}
&&L(U_{i}+U_0)=i; \ i\geq 1,
\\
&&U_0=\{(2,3),(4,5),...,(2n-2,2n-1),(2n,1)\},
\end{eqnarray}
and the recurrence relation between $U_i$ and $U_{i+1}$ is

\begin{eqnarray}
U_{i+1}=U_i-\{(a_i,b_i),(d_i,c_i)\}+\{(a_i,c_i),(d_i,b_i)\}; \ i\geq 1,
\end{eqnarray}
with its first term

\begin{eqnarray}
U_1=\{(1,2),(3,4),...,(2n-1,2n)\}.
\end{eqnarray}
$U=\{(a,b),(b,c),...\}$ representing that the element $a$ is connected with $b$, $b$ is connected with $c$..., is a set of 2-dimensional row matrices $(i,j)$. $L(U)$ is a function of $U$, which equals the number of loops of the elements in $U$. For example, $U=\{(1,6),(2,3),(3,4),(4,5),(5,2),(6,1)\}$, which consists of two loops $1-6-1$ and $2-3-4-5-2$, therefore $L(U)=2$.

In the above expression, $U_i+U_0$ represents the configuration, which contains $i$ fermion loops, before the $i$th transition, then one can easily see that the sum condition $L(U_{i+1}+U_0)=i+1$ ensures that only transitions to states with one more fermion loop, as well as one more order of $N_c$, are allowed, while the recurrence relation between $U_i$ and $U_{i+1}$ shows how the configurations of these states translate.

One can use the above formula to get any 2n-point correlators, as well as the 6-point one, which can be subsequently derived as an illustration. For $n=3$, one gets

\begin{eqnarray}
&&[e^{C_F\mu^2A_1}
        +\sum_{a_1,b_1,c_1,d_1,A_1,A_2}F_{a_1b_1c_1d_1}
        \left[\frac{e^{C_F\mu^2A_1}}{(A_1-A_2)}+\frac{e^{C_F\mu^2A_2}}{(A_2-A_1)} \right]
        +\sum_{ \substack{a_1,b_1,c_1,d_1,A_1,A_2,\\a_2,b_2,c_2,d_2,A_3} }F_{a_1b_1c_1d_1}F_{a_2b_2c_2d_2}\nonumber
\\
&&\times\left[ \frac{e^{C_F\mu^2A_1}}{(A_1-A_2)(A_1-A_3)}+\frac{e^{C_F\mu^2A_2}}{(A_2-A_1)(A_2-A_3)}+\frac{e^{C_F\mu^2A_3}}{(A_3-A_1)(A_3-A_2)}\right]
  ]T,\label{6}
\end{eqnarray}
where

\begin{eqnarray}
&&U_1=\{(1,2),(3,4),(5,6)\};A_1=L_{12}+L_{34}+L_{56},
\end{eqnarray}
for the second term

\begin{eqnarray}
&(a_1,b_1,d_1,c_1,U_2,A_2)=&(1,2,3,4,\{(1,4),(3,2),(5,6)\},L_{14}+L_{32}+L_{56}),\nonumber
\\
&&(1,2,5,6,\{(1,6),(3,4),(5,2)\},L_{16}+L_{34}+L_{52}),\nonumber
\\
&&(3,4,5,6,\{(1,2),(3,6),(5,4)\},L_{12}+L_{36}+L_{54}),
\end{eqnarray}
for the third term

\begin{eqnarray}
&&(a_1,b_1,d_1,c_1,U_2,A_2,a_2,b_2,d_2,c_2,U_3,A_3)\nonumber
\\
&=&(1,2,3,4,\{(1,4),(3,2),(5,6)\},L_{14}+L_{32}+L_{56},1,4,5,6,\{(1,6),(3,2),(5,4)\}),\nonumber
\\
&& (1,2,5,6,\{(1,6),(3,4),(5,2)\},L_{16}+L_{34}+L_{52},3,4,5,2,\{(1,6),(3,2),(5,4)\}),\nonumber
\\
&& (3,4,5,6,\{(1,2),(3,6),(5,4)\},L_{12}+L_{36}+L_{54},1,2,3,6,\{(1,6),(3,2),(5,4)\}).
\end{eqnarray}

Putting the values of the above sum variables and the tadpole term $T$ into Eq. (\ref{6}), one can get the exact 6-point correlator shown in Eq. (\ref{6-point}).

When calculating the multiple scattering factors in the multi-particle production processes, one may encounter the case that several coordinates coincide in the multipole amplitude, which can be entirely derived by taking the corresponding limits of coordinates in the above general expression of 2n-point function. We derive the simplest case with an explicit expression that only two different coordinates $x_1$ and $x_2$ are involved in correlators as an example. Taking the limit $x_{3,4},x_{5,6}\rightarrow x_{1,2}$, namely $A_{1,2,3}\rightarrow3L_{1,2}$, in Eq. (\ref{6}), which gives

\begin{eqnarray}
\frac{1}{N_c}\left\langle\text{Tr}\left[ U_{1}U_{2}^{\dagger}U_{1}U_{2}^{\dagger}U_{1}U_{2}^{\dagger}\right]\right\rangle=e^{-3\Gamma_{12}}\left[1+3C_F\mu^2F_{1221}+\frac{3}{2}(C_F\mu^2F_{1221})^2\right].
\end{eqnarray}
In the same way, one also gets

\begin{eqnarray}
\frac{1}{N_c}\left\langle\text{Tr}\left[ U_{1}U_{2}^{\dagger}U_{1}U_{2}^{\dagger}\right]\right\rangle=e^{-2\Gamma_{12}}\left[1+C_F\mu^2F_{1221}\right],
\end{eqnarray}
and

\begin{eqnarray}
\frac{1}{N_c}\left\langle\text{Tr}\left[U_{1}U_{2}^{\dagger}U_{1}U_{2}^{\dagger}U_{1}U_{2}^{\dagger}U_{1}U_{2}^{\dagger}\right]\right\rangle
=e^{-4\Gamma_{12}}\left[1+6C_F\mu^2F_{1221}+\frac{16}{2!}(C_F\mu^2F_{1221})^2+\frac{16}{3!}(C_F\mu^2F_{1221})^3\right].\label{8simp}
\end{eqnarray}
And the expression for the 2n-point correlator with two different coordinates involved can be also obtained as 

\begin{eqnarray}
\frac{1}{N_c}\left\langle\text{Tr}\left[(U_{1}U_{2}^{\dagger})^{n}\right]\right\rangle
=e^{-n\Gamma_{12}}\sum_{i=0}^{n-1}\left[\frac{n^{i-1}C_n^{i+1}}{i!}(C_F\mu^2F_{1221})^i\right],
\end{eqnarray}
where the factor $n^{i-1}C_n^{i+1}$ is the exact number of color transition ways from the initial color structure with one fermion loop to the color singlet state with $i+1$ fermion loops after the $i$th transition. For the octupole case, namely $n=4$, we have $C_n^2=6$, $nC_n^3=n^2C_n^4=16$, which are the exact corresponding coefficients in the second, the third and the fourth term in Eq. (\ref{8simp}), and also equal the numbers of the terms with one, two and three factors $F_{abcd}$ in Eq. (\ref{8result}), respectively.

The general expression contains different terms corresponding to the number of factors $F_{a_ib_ic_id_i}$, which represents the number of transitions between the color singlet states. One can recognise that the 8-point correlator has one term with no color transitions, and several terms with one, two and three transitions, while it is not difficult to see from Eq. (\ref{2n-point}) that the largest number of color transitions allowed in the terms of 2n-point correlators is $n-1$.

\section{Conclusion}
In conclusion, we find that, in the large-$N_c$ limit in the MV model, the octupole amplitude as well as general 2n-point correlators can be written in analytical forms, which may help us to outline the underlying dynamics and estimate the size of finite-$N_c$ corrections to multi-particle productions, and provide initial conditions for small-$x$ evolutions.

\begin{acknowledgments}
We thank Dr. Bo-Wen Xiao for discussions and comments. This material is based upon work supported by the Natural Science Foundation of China (NSFC) under Grant No. 11575070, No. 11435004 and by the Ministry of Science and Technology of China under Projects No. 2014CB845404.
\end{acknowledgments}

\end{document}